\documentclass[runningheads,a4paper]{llncs}

\usepackage[english]{babel}
\usepackage[utf8]{inputenc}
\usepackage{amsmath}
\usepackage{graphicx}
\usepackage[colorinlistoftodos]{todonotes}
\usepackage{xspace}
\usepackage{comment}

\usepackage{amssymb,amsfonts,textcomp,setspace,multirow,tabu}
\usepackage{stmaryrd}
\usepackage{tikz}

\usepackage[vlined,ruled]{algorithm2e}

\newcommand{\SV}[1]{#1}
\newcommand{\LV}[1]{}

\SV{\excludecomment{pf}}

\LV{\newenvironment{pf}
{\begin{proof}}
{\hspace*{\fill}\framebox{}\end{proof}}
}
\newenvironment{prf}
{\begin{proof}}
{\hspace*{\fill}\framebox{}\end{proof}}

 \newcommand{\opt}{\operatorname{opt}}
 
\newcommand{\MCC}{\textsc{Multicoloured Clique}\xspace}
\newcommand{\MD}{\textsc{Minimum Domination}\xspace}

\newcommand{\UD}{\textsc{Upper Domination}\xspace}
\newcommand{\UIr}{\textsc{Upper Irredundance}\xspace}

\newcommand{\CUD}{\textsc{Co-Upper Domination}\xspace}
\newcommand{\MIS}{\textsc{Maximum Independent Set}\xspace}

\newcommand{\MVC}{\textsc{Minimum Vertex Cover}\xspace}

\title{Algorithmic Aspects of Upper Domination}

\author{C. 
Bazgan\inst{1,2}
\and L. 
Brankovic\inst{3,4}
\and  K. 
Casel\inst4
\and H. 
Fernau\inst4
\and K. 
Jansen\inst5
\and M. 
Lampis\inst{2}
\and M. 
Liedloff\inst{6}
\and J. 
Monnot\inst{2}
\and V. Th. 
Paschos\inst{1,2}
}

\authorrunning{Bazgan \emph{et al.}}

\institute{Institut Universitaire de France\\
\and 
PSL, University of Paris-Dauphine, CNRS, LAMSADE UMR 7243, \\ F-75775 Paris Cedex 16, France\\
\email{bazgan@lamsade.dauphine.fr},
\email{michail.lampis@lamsade.dauphine.fr}, 
\email{jerome.monnot@dauphine.fr}, \email{paschos@lamsade.dauphine.fr}
\and \LV{School of Electrical Engineering and Computer Science,} The University of Newcastle, Callaghan, NSW 2308, Australia\\
\email{ljiljana.brankovic@newcastle.edu.au}
\and 
Universit\"at Trier, Fachbereich 4,
Informatikwissenschaften,\\ 
D-54286 Trier, Germany\SV{,}\LV{\\}
\email{casel,fernau@uni-trier.de}
\and
Universität Kiel, Institut f\"ur Informatik\SV{,}\LV{\\}
D-24098 Kiel, Germany\SV{,}\LV{\\}
\email{kj@informatik.uni-kiel.de}
\and 
Univ. Orl\'eans, INSA Centre Val de Loire, LIFO EA 4022,\\ F-45067 Orl\'eans, France\SV{,}\LV{\\}
\email{mathieu.liedloff@univ-orleans.fr}
}

\date{\today}
\begin{document}
\maketitle

\begin{abstract}
In this paper we study combinatorial and algorithmic resp. complexity questions of upper domination, i.e., the maximum cardinality of a minimal dominating set in a graph.
We give a full classification of the 
related maximisation and minimisation problems, as well as the related parameterised problems, on general graphs and on graphs of bounded degree, and we also study planar graphs.
\end{abstract}

\section{Introduction}


The famous domination chain 
$$\textrm{ir}(G)\leq \gamma(G)\leq 
i(G)\leq \alpha(G)\leq \Gamma(G)\leq \textrm{IR}(G)
$$
links 
parameters related to the fundamental notions of independence, domination and irredundance in graphs; see \cite{HHS98}. 
Out of the six parameters, the least  studied appears to be the upper domination parameter $\Gamma(G)$. This paper is devoted to changing this impression. 

\LV{\subsection{Basic notions}}\SV{\paragraph{Basic notions.}}
Throughout this paper, we only deal
with undirected simple graphs $G=(V,E)$. 
In the following, we explain the main graph theory notions  that we use in this paper and refer the reader to Appendix~\ref{basic_notions} and any graph theory textbook  for other standard concepts and notations. 
\LV{
The number of vertices $|V|$ is also known as the order of $G$. 
As usual, $N(v)$ denotes the open neighbourhood of $v$ in a graph $G$, and $N[v]$ is the closed neighbourhood of $v$ in $G$, i.e., $N[v]=N(v)\cup\{v\}$. These notions can be easily extended to vertex sets $X$, e.g., $N(X)=\bigcup_{x\in X}N(x)$. 
The cardinality of $N(v)$ is also known as the degree of $v$,
denoted as $deg(v)$. The maximum degree in a graph is usually written
as $\Delta$. A graph of maximum degree three is called subcubic, and if actually all degrees equal three, it is called a cubic graph.}

 Given a  graph $G=(V,E)$, a subset $S$ of $V$ is a \emph{dominating set} if every vertex $v\in V\setminus S$ has at least one neighbour in $S$, i.e., if $N[S]=V$. A dominating set is minimal if no proper subset of it is a dominating set.
\LV{Likewise, a vertex set $I$ is \emph{independent} if $N(I)\cap I=\emptyset$. An independent set is maximal if no proper superset
 is independent.}
 In the following we use classical notations: $\gamma(G)$ and $\Gamma(G)$ are the minimum and maximum cardinalities over all minimal dominating sets in $G$, $\alpha(G)$ is the maximum cardinality of an independent set, $i(G)$ is the minimum cardinality 
 of a maximal independent set, and 
 $\tau(G)$ is the size of a  minimum vertex cover, which equals $|V|-\alpha(G)$ by Gallai's identity.
 A minimal dominating set $D$ of $G$ with $|D|=\Gamma(G)$ is also known as an \emph{upper dominating set} of $G$.

For any subset $S\subseteq V$ and $v\in S$ we define the private neighbourhood of $v$ with respect to $S$ as $pn(v,S):=N[v]-N[S-\{v\}]$. Any $w\in pn(v,S)$ is called a \emph{private neighbour of $v$ with respect to $S$}. If the set $S$ is clear from the context, we will omit the ``with respect to $S$" part.
$S$ is called \emph{irredundant} if every vertex in $S$ has at least one private neighbour, i.e., if $|pn(v,S)|>0$ for every $v\in S$.
$\textrm{IR}(G)$
denotes the cardinality of the largest irredundant set in $G$, while 
$\textrm{ir}(G)$ is the cardinality of the smallest maximal irredundant set in $G$. 
 
 The domination chain is largely due to the following two combinatorial properties:
 (1)  Every maximal independent set is a minimal dominating set. 
 (2)
A dominating set $S\subseteq V$ is minimal if and only if $|pn(v,S)|>0$ for every $v\in S$. Observe that $v$ can be a private neighbour of itself, i.e., a dominating set is minimal if and only if it is also an irredundant set. 
Actually, every minimal dominating set is also a maximal irredundant set.

\LV{\subsection{Our combinatorial problems}}\SV{\paragraph{Our combinatorial problems.}}
We will mostly deal with the following two combinatorial problems, investigating algorithmic and complexity aspects from different angles, for instance, approximation and parameterised complexity.

\smallskip
\noindent
\fbox{\begin{minipage}{.47\textwidth}
\UD\nopagebreak\\\nopagebreak
{\bf Input:} A   graph $G=(V,E)$, a non-negative integer $k$.\\\nopagebreak
{\bf Question:}  Is $\Gamma(G) \geq k$?
\end{minipage}}\hfill  
\fbox{\begin{minipage}{.47\textwidth}
\noindent\CUD\\\nopagebreak
{\bf Input:} A   graph $G=(V,E)$, a non-negative integer $\ell$. \\\nopagebreak
{\bf Question:}   Is $\Gamma(G) \geq |V|-\ell$?
\end{minipage}}
\smallskip

From the perspective of classical complexity theory, both problems are trivially equivalent and known to be NP-complete for quite some time~\cite{CheFHJ90}. 
Slightly abusing notation, we will also consider them from 
the perspective of parameterised complexity. In that case,
$k$ and $\ell$ turn out to be  the natural parameters of these 
problems, which turn both problems into dual problems in the
parameterised complexity sense of this word.
Actually, as we mostly consider this natural parameterisation,
no further mentioning of the choice of the parameter is necessary
when putting down the results. 
Finally, we also consider these problems from the perspective
of optimisation, again slightly abusing notations. \UD is then 
a maximisation problem, while \CUD is a minimisation problem.

\LV{\subsection{On some complexity notions}}

\paragraph{Parameterised Complexity.}
We mainly refer to the newest textbook \cite{DowFel2013} in the area.
Important notions that we will make use of include the parameterised complexity 
classes FPT, W[1] and W[2], parameterised reductions and kernelisation.
In this area, it has also become customary not only to suppress constants (as in the well-known $O$ notation), but also even polynomial-factors, leading to the so-called $O^*$-notation.

\paragraph{Approximation.}
We mostly refer to the textbook~\cite{Ausetal99}. 
Given an optimisation problem and an instance
$I$ of this problem, we use $|I|$ to denote the size of $I$,
$\opt(I)$ to denote the optimum value of $I$, and
$val(I,S)$ to denote the value of a feasible solution $S$ of
instance $I$. The {\em performance ratio\/} of $S$ (or {\it
approximation factor}) is
$r(I,S)=\max\left\{\frac{val(I,S)}{\opt(I)},
\frac{\opt(I)}{val(I,S)}\right\}.$ The  {\em error} of $S$,
$\varepsilon(I,S)$, is defined by $\varepsilon(I,S)= r(I,S)-1.$ For a function $f$, an algorithm  is an {\it
$f(|I|)$-approximation\/}, if for every instance $I$ of the problem,
it returns a solution $S$ such that $r(I,S) \leq f(|I|).$  When $f$ is $1+\varepsilon$ for any $\varepsilon >0$ and the algorithm runs in polynomial time in $|I|$ and in exponential time in $\frac 1\varepsilon$, the problem admits a polynomial time approximation scheme (PTAS for short). APX is the class of optimisation problems for which there exists a polynomial time $c$-approximation algorithm for some constant $c > 1$. 

For providing hardness proofs in the area of approximation algorithms, $L$-reductions have become a kind of standard. An optimisation problem which is APX-hard under $L$-reduction has no polynomial-time approximation scheme if P $\neq$ NP.
We will deviate a bit from the $L$-reduction methodology, using $E$-reductions instead in one place.

\LV{The notion of an $E$-reduction ({\it error-preserving}
reduction) was introduced   by Khanna et al. \cite{Khaetal98}. A
problem $A$ is called {\it $E$-reducible} to a problem $B$,
if there exist polynomial 
time computable functions $f$, $g$ and a
constant $\beta$ such that
\begin{itemize}
\item $f$ maps an instance $I$ of $A$ to an instance $I'$ of $B$
such that $\opt(I)$ and $\opt(I')$ are related by a polynomial
factor, i.e. there exists a polynomial $p$ such that
$\opt(I')\leq p(|I|) \opt(I)$, \item $g$ maps any solution $S'$ of $I'$
to one solution $S$ of $I$ such that $\varepsilon(I,S)\leq \beta
\varepsilon(I',S')$.
\end{itemize}

An important property of an  $E$-reduction is that it can be applied
uniformly to all levels of approximability; that is, if $A$ is
$E$-reducible to $B$ and $B$ belongs to $\cal{C}$ then $A$
belongs to $\cal{C}$ as well, where $\cal{C}$ is a class of
optimisation problems with any kind of approximation guarantee
(see also \cite{Khaetal98}).}

\LV{\subsection{Our main results}}\SV{\paragraph{Main results.}}
(1) We link minimal dominating sets to a decomposition of the vertex set that turns out to be a crucial tool for deriving our combinatorial and computational results.
(2) We bound the upper domination number by the maximum independence number, the order and also by the maximum degree of a graph.
(3) We explain the particular hardness of dealing with this graph parameter by showing NP-hardness of an extension problem variant that rules out certain 
natural greedy strategies for filling up partial solutions.
(4) We show that \UD is W[1]-hard, so very likely not to belong to FPT. (5) Conversely, \CUD is in FPT, which we prove by providing both a kernelisation and a branching algorithm.
(6) Likewise, \CUD is in APX, while
\UD is not $n^{1-\varepsilon}$-approximable for any $\varepsilon>0$, unless P=NP.
(7) \UD is NP-hard even on cubic graphs. 
(8) Both \UD and \CUD are APX-complete on bounded degree graphs.
(9) \UD is in FPT for graphs of bounded degree.

\SV{For reasons of space, proofs and other  details were moved into an appendix to this extended abstract.}


\section{\LV{Notes on the combinatorial}\SV{On the} structure of minimal dominating sets}\label{sec-FIPO}

Any minimal dominating set $D$ for a graph $G=(V,E)$ can be associated with a partition of the set of vertices $V$ into four sets $F,I,P,O$ 
given by: $I:=\{v\in D\colon v\in pn(v,D)\}$, $F:=D-I$, 
$P\in\{B\subseteq N(F)\cap(V-D)\colon |pn(v,D)\cap B|=1 $ for all $ v\in F\}$ with $|F|=|P|$, $O=V-(D\cup P)$. 
This representation is not necessarily unique since there might be different choices for the sets $P$ and $O$, but for every partition of this kind, the following properties hold:
\begin{enumerate}
\item Every vertex $v\in F$ has at least one neighbour in $F$, called a \textbf{f}riend.
\item The set $I$ is an \textbf{i}ndependent set in $G$.
\item The subgraph induced by the vertices $F\cup P$ has an edge cut set separating $F$ and $P$ that is, at the same time, a perfect matching; hence, $P$ can serve as the 
set of \textbf{p}rivate neighbours for~$F$. 
\item The neighbourhood of a vertex in $I$ is always a subset of $O$, which are otherwise the \textbf{o}utsiders.
\end{enumerate}

\LV{\todo[inline]{HF: Maybe, we could add a small picture for illustration in the long version.}}
 
 This partition is also related to a different characterisation of $\Gamma(G)$ in terms of so-called upper perfect neighbourhoods~\cite{HHS98}.

\begin{lemma} \label{|I|<=alpha-2}
For any connected graph $G$ with $n>0$ vertices and an upper dominating set $D$ with  an associated partition $(F,I,P,O)$ as defined above, if  $|D| = \Gamma(G) > \alpha(G)$ then $|I| \leq \alpha(G) -2$.
\end{lemma}

\newcommand{\proofofLemmaalphatwo}{Let $G$ be a connected graph with $n>0$ vertices and let $D$ be an upper dominating set  with  an associated partition $(F,I,P,O)$ as defined above.

We first show that if $\Gamma(G) > \alpha(G)$ then $|F| \geq 2$ (in fact, one can show that then $|F| \geq 3$ but that is not necessary for our proof). Indeed, if $|F|=0$, then the upper dominating set is also an independent set, and thus $\Gamma(G) = \alpha(G)$, and according to our definition of partition $(F,I,P,O)$, we have $|F| \ne 1$ (see Property 1 of this partition). 

Now, if $|F| \geq 2$ then the subgraph of $G$ induced by $F \cup P$ contains an independent set of size $2$ consisting of a vertex in $F$, say $v$, and a vertex in $P$, say $u$, such that $v$ and $u$ are not adjacent. Since in the original graph $G$, there are no edges between the vertices in $I$ and the vertices in $F \cup P$ (Property 4), $I \cup \{u,v\}$ forms an independent set of size $|I| +2$. This sets a lower bound on the independence number and we have  $\alpha(G) \geq |I| +2$, that is, $|I| \leq \alpha(G) -2$. 

From the above, it follows that if $\Gamma(G) > \alpha(G)$ then $|I| \leq \alpha(G) -2$.}
\begin{pf}\proofofLemmaalphatwo
\end{pf}

\begin{lemma} \label{bounds_on_Gama}
For any connected graph $G$ with $n>0$ vertices we have: 
\begin{equation}\alpha(G)\ \leq \ \Gamma(G)\ \leq \ \ \max \  \left\{\alpha(G), \frac{n}{2} + \frac{\alpha(G)}{2}-1\right\} 
\label{upperdom_is}
\end{equation}
\end{lemma}

\newcommand{\proofofLemmaboundsonGama}{We consider a graph $G$ with $n>0$ vertices and an upper dominating set $D$ with  an associated partition $(F,I,P,O)$ as defined above. The left inequality comes from the fact that any maximal independent set is a minimal dominating set. 
 For the right inequality, we examine separately the following two cases.
 \begin{enumerate}
 \item $\Gamma(G) = \alpha(G)$. Then we trivially have $\Gamma(G) \leq \alpha(G)$.
 \item $\Gamma(G) > \alpha(G)$.
 
From the  fact that $|F|=|P|$ (from Property 3) we have $|F| =  \frac {n-|I|-|O|}{2}  
\leq \left\lfloor \frac {n-|I|}{2} \right\rfloor $ and thus 
$$ \Gamma(G) = |F| + |I| \leq \left\lfloor  \frac {n+|I|}{2} \right\rfloor $$

From the above and Lemma \ref{|I|<=alpha-2} we have 
$$ \Gamma(G) \   \leq \ \left\lfloor  \frac {n+|I|}{2} \right\rfloor\ \leq \ \left\lfloor \frac  {n+\alpha(G) -2}{2}\right \rfloor \ \leq \ \frac{n}{2} + \frac{\alpha(G)}{2}-1 $$

\end{enumerate}

This concludes the proof of the claim.}
%
%

\begin{pf}\proofofLemmaboundsonGama
\end{pf}

The following lemma generalises the earlier result on upper bounds on $\textrm{IR}(G)$ (and hence on $\Gamma(G)$) for
$\Delta$-regular graphs $G$, which is 
$\textrm{IR}(G)\leq n/2$; see~\cite[Proposition~12]{HenSla96}. Notice that our result generalises the mentioned result of Henning and Slator, as minimum and maximum degrees coincide for regular graphs.

\begin{lemma}\label{bounds_on_Gama_with_Delta}
For any connected graph $G$ with $n>0$ vertices, minimum degree $\delta$ and maximum degree $\Delta$, we have: 
\begin{equation}\alpha(G) \ \leq \ \Gamma(G)\ \leq  \ \max \ \left\{\alpha(G),  \ \frac{n}{2} + \frac{\alpha(G)(\Delta-\delta)}{2\Delta}-\frac {\Delta-\delta}{\Delta}\right\} 
\label{upperdomdelta_is}
\end{equation}
\end{lemma}

\newcommand{\proofofLemmaboundsonGamawithDelta}{Let $G$ be a connected graph with $n>0$ vertices,  maximum degree $\Delta$ and an upper dominating set $D$ with  an associated partition $(F,I,P,O)$ as defined above. Our argument is similar to the one in  Lemma \ref{bounds_on_Gama}: The left inequality comes from the fact that any maximal independent set is a minimal dominating set. 
For the right inequality, we examine separately the following two cases.
 \begin{enumerate}
 \item $\Gamma(G) = \alpha(G)$. 
 Then we trivially have $\Gamma(G) \leq \alpha(G)$.
 \item $\Gamma(G) > \alpha(G)$.
 Again, we obtain:
$$ \Gamma(G) = |F| + |I| =   \frac {n+|I|-|O|}{2}  $$

We next  derive an improved lower bound on $|O|$. Let $e$ be the number of edges adjacent with vertices from $I$. As $G$ is of minimum degree $\delta$,  we have  $e \geq \delta|I|$. As the  vertices in $I$ are only adjacent with the vertices in $O$, there are at least $e$ edges that have exactly one end vertex in $O$. Since $G$ has maximum degree $\Delta$, we have that $|O| \geq \left\lceil \frac{e}{\Delta} \right\rceil \geq  \left\lceil \frac{\delta|I|}{\Delta} \right\rceil$.

From the above and Lemma \ref{|I|<=alpha-2} we have 
\begin{eqnarray*}\Gamma(G)   \  
\leq & \left\lfloor  \frac {n+|I|- \left\lceil \frac {\delta|I|}{\Delta} \right\rceil }{2} \right\rfloor &  
\leq \  \frac {n+|I|- \frac {\delta|I|}{\Delta} }{2}
= \frac {n+ \frac {(\Delta-\delta)|I|}{\Delta} }{2}\\
\leq & \frac {n+ \frac {(\Delta-\delta)}{\Delta}(\alpha(G)-2) }{2} &
=  \ \frac{n}{2} + \frac{\Delta-\delta}{2\Delta}\alpha(G)-\frac{\Delta-\delta}{\Delta}
\end{eqnarray*}
\end{enumerate}

To combine these two upper bounds we take their maximum, which concludes the proof.}
\begin{pf}\proofofLemmaboundsonGamawithDelta
\end{pf} 

 Given a graph on $n$ vertices, we subtract $n$ in  the inequalities (\ref{upperdom_is})\SV{; thus:}\LV{ and thus $$ \frac{n}{2} - \frac{\alpha(G)}{2}+1\leq n- \Gamma(G)\leq n-\alpha(G) $$
 so that we obtain the following relationship for the vertex cover number $\tau(G)$.}
 
 \begin{lemma}\label{lem-complupperdom}
 Let $G$ be a connected graph of order $n$. Then, 
  \begin{equation}
 \frac{\tau(G)}{2}+1 \leq n- \Gamma(G)\leq \tau(G)
 \label{complupperdom}
\end{equation}
\end{lemma}

\LV{Observe 
that all our  bounds derived in this section are also valid for the upper irredundance number $\textrm{IR}(G)$ (instead of $\Gamma(G)$).

We also like to point the reader to \cite{Zve2003}, where another interesting combinatorial bound was shown, namely
$$\textrm{IR}(G)-\alpha(G)\leq \left\lceil\frac{\Delta-2}{2\Delta}n \right\rceil
$$
This somehow bounds the difference between the two entities that are maximised in the previous lemma (recall that the same bounds hold for upper irredundance and for upper domination).

As our hardness results will also prove hardness for cubic triangle-free graphs, the following bound \cite[Theorem 5]{Zve2003}
could be also interesting; in that case,
$\alpha(G)$\todo{to be continued ...}
}

\section{What makes this problem that hard?}
\label{sec-extension}

Algorithms working on combinatorial graph problems often try to look at local parts of the graph and then extend some part of the (final) solution that was found and fixed so far. This type of strategy is at least difficult to implement for \UD, as the following example shows.

First, consider a graph $G_n$ that consists of two cliques with vertices
$V_n=\{v_1,\dots,v_n\}$ and $W_n=\{w_1,\dots,w_n\}$, where the only edges connecting both
cliques are $v_iw_i$ for $1\leq i\leq n$.
Observe that $G_n$ has as minimal dominating sets $V_n$, $W_n$, and
$\{v_i,w_j\}$ for all $1\leq i,j\leq n$.
For $n\geq 3$, the first two are upper dominating sets, while the last $n^2$ many are minimum dominating sets.
If we now add a vertex $v_0$ to $G_n$, arriving at graph $G_n'$,
and make $v_0$ adjacent to all vertices in $V_n$, then $V_n$
is still a minimal dominating set, but $W_n$ is no longer a dominating set. Now, we have $\{v_i,w_j\}$ for all $0\leq i\leq n$ and all $1\leq j\leq n$ as minimum dominating sets.
But, if we add one further vertex, $w_0$ to $G_n'$ to obtain $G_n''$
and  make $w_0$ adjacent to all vertices in $W_n$, then
all upper dominating sets are also minimum dominating sets and vice versa.
 This shows that we cannot consider vertices one by one, but must rather 
look at the whole graph.

For many maximisation problems, like \UIr or \MIS,  
it is trivial to obtain a feasible solution that extends a given vertex set, namely by some greedy strategy, or to know that no such extension exists. This is a different situation with \UD, as we see next.
To be able to reason about this problem, let us first 
define it formally.

\begin{center}
\fbox{\begin{minipage}{.95\textwidth}
\noindent{\sc Minimal Dominating Set Extension}\\\nopagebreak
{\bf Input:} A   graph $G=(V,E)$, a set $S \subseteq V$. \\\nopagebreak
{\bf Question:}  Does $G$ have a minimal dominating set $S'$ with $S'\supseteq S$?
\end{minipage}}
\end{center}

Notice that this problem is trivial on some input with $S=\emptyset$\SV{ by using a greedy approach}.
\LV{Namely, in that case, we can start a greedy algorithm from $D:=V$, 
gradually deleting vertices from $D$ until the domination property would be destroyed. So, we end up with a set $D$ from which we cannot remove any vertex while keeping the domination property of $D$.} However, this strategy might fail if the given set $S$ is bigger, as we must also maintain the property of being superset of $S$. This difficulty is very hard to overcome, as the next result shows.
Its proof is based on a reduction from {\sc 4-Bounded Planar 3-Connected SAT} (4P3C3SAT) \cite{Kra94}.

\begin{theorem}\label{thm-MDSE-hardness}
{\sc Minimal Dominating Set Extension} is NP-complete, even when restricted to planar cubic graphs.
\end{theorem}

\newcommand{\proofofTheoremMDSEhardness}{Membership in NP is obvious. NP-hardness can be shown by reduction from {\sc 4-Bounded Planar 3-Connected SAT} (4P3C3SAT) \cite{Kra94}:  
Consider an instance of 4P3C3SAT with clauses $c_1,\dots, c_m$ and variables $v_1,\dots, v_n$. By definition, the graph $G=(V,E)$ with $V=\{c_1,\dots,c_m\}\cup\{v_1,\dots,v_n\}$ and $E=\{(c_j,v_i)\colon v_i$ or $ \bar v_i$ is literal of $c_j\}$ is planar. Replace every vertex $v_i$ by six new vertices $f_i^1,x_i^1,t_i^1,t_i^2,x_i^2,f^2_i$ with edges $(f_i^j,x^j_i),(t_i^j,x_i^j)$ for $j=1,2$. 
If $v_i$  (positive) is a literal in more than two clauses, add the edge $(f^1_i,f^2_i)$, else add the edge $(t^1_i,t^2_i)$.  By definition of the problem 4P3C3SAT, each variable appears in at most four clauses and this procedure of replacing the variable-vertices in $G$ by a $P_6$ preserves planarity. To see this, consider any fixed planar embedding of $G$ and any variable $v_i$ which appears in clauses $c_1,c_2,c_3,c_4$, in the embedding placed like in the picture below:
 \begin{center}
\begin{tikzpicture}
\tikzstyle{every node}=[inner sep=0.5mm,draw,circle] 
	\node[draw=none, fill=none] (n) at (-4.2,1.2) {{\small $v_i$}};	
	\node (vi) at (-4,1) {};
	\node[draw=none, fill=none] (n) at (-4,2.3) {{\small $c_1$}};		
	\node (l1) at (-4,2) {};
	\node[draw=none, fill=none] (n) at (-4,-0.3) {{\small $c_2$}};	
	\node (l2) at (-4,0) {};
	\node[draw=none, fill=none] (n) at (-5,1.3) {{\small $c_3$}};	
	\node (l3) at (-5,1) {};
	\node[draw=none, fill=none] (n) at (-3,1.3) {{\small $c_4$}};		
	\node (l4) at (-3,1) {};
  \foreach \from/\to in {l1/vi,l2/vi,l3/vi,l4/vi}
    \draw (\from) -- (\to);
\end{tikzpicture}
\end{center} 
Depending on whether $v_i$ appears negated or non-negated in these clauses, we differentiate between the following cases; in the following pictures, vertices plotted in black are the ones to be put into the vertex set $S$ predetermined to be in the minimal dominating set.\\

If $v_i$ is literal in $c_1,c_2,c_3$ and $\bar v_i$ literal in $c_4$:
 \medskip
 \begin{center}
\begin{tikzpicture}
\tikzstyle{every node}=[inner sep=0.5mm,draw,circle] 

	\node[draw=none, fill=none] (n) at (1.3,0) {{\small $t_i^1$}};			
	\node (t1) at (1,0) {};
	\node[draw=none, fill=none] (n) at (1.3,-0.5) {{\small $c_2$}};			
	\node (c2) at (1,-0.50) {};
	\node[draw=none, fill=none] (n) at (-0.5,1.3) {{\small $c_3$}};	
	\node (c3) at (-0.5,1) {};		
	\node[draw=none, fill=none] (n) at (1,1.3) {{\small $f_i^1$}};	
	\node (f1) at (1,1) {};	
	\node[draw=none, fill=none] (n) at (2.5,1.25) {{\small $c_4$}};
	\node (c4) at (2.5,1) {};		
	\node[draw=none, fill=none] (n) at (2,1.3) {{\small $f_i^2$}};
	\node (f2) at (2,1) {};	
	\node[draw=none, fill=none] (n) at (0.7,2.5) {{\small $c_1$}};	
	\node (c1) at (1,2.5) {};	
	\node[draw=none, fill=none] (n) at (0.7,2) {{\small $t_i^2$}};	
	\node (t2) at (1,2) {};	
	\node[draw=none, fill=none] (n) at (1.3,0.5) {{\small $ x_i^1$}};	
	\node[fill=black] (x1) at (1,0.5) {};	
	\node[draw=none, fill=none] (n) at (1.7,1.7) {{\small $ x_i^2$}};		
	\node[fill=black] (x2) at (1.5,1.5) {};	
				
  \foreach \from/\to in {f1/x1,t1/x1,f2/x2,t2/x2, f1/f2,c3/t1,c2/t1,f2/c4,t2/c1}
    \draw (\from) -- (\to);
\end{tikzpicture}
\end{center} 
  \medskip

If $v_i$ is literal in $c_2,c_4$ and $\bar v_i$ literal in $c_1,c_3$:
 \medskip
 \begin{center}
\begin{tikzpicture}
\tikzstyle{every node}=[inner sep=0.5mm,draw,circle] 

	\node[draw=none, fill=none] (n) at (1.3,0) {{\small $t_i^1$}};			
	\node (t1) at (1,0) {};
	\node[draw=none, fill=none] (n) at (1.3,-0.5) {{\small $c_2$}};			
	\node (c2) at (1,-0.50) {};
	\node[draw=none, fill=none] (n) at (-0.5,1.3) {{\small $c_3$}};	
	\node (c3) at (-0.5,1) {};		
	\node[draw=none, fill=none] (n) at (0,1.3) {{\small $f_i^1$}};	
	\node (f1) at (0,1) {};	
	\node[draw=none, fill=none] (n) at (2.5,1.25) {{\small $c_4$}};
	\node (c4) at (2.5,1) {};		
	\node[draw=none, fill=none] (n) at (2,1.3) {{\small $t_i^2$}};
	\node (t2) at (2,1) {};	
	\node[draw=none, fill=none] (n) at (0.7,2.5) {{\small $c_1$}};	
	\node (c1) at (1,2.5) {};	
	\node[draw=none, fill=none] (n) at (0.7,2) {{\small $f_i^2$}};	
	\node (f2) at (1,2) {};	
	\node[draw=none, fill=none] (n) at (0.25,0.25) {{\small $ x_i^1$}};	
	\node[fill=black] (x1) at (0.5,0.5) {};	
	\node[draw=none, fill=none] (n) at (1.7,1.7) {{\small $ x_i^2$}};		
	\node[fill=black] (x2) at (1.5,1.5) {};	
				
  \foreach \from/\to in {f1/x1,t1/x1,f2/x2,t2/x2, t1/t2,c3/f1,c2/t1,t2/c4,f2/c1}
    \draw (\from) -- (\to);
\end{tikzpicture}
\end{center} 
  \medskip

If $v_i$ is literal in $c_1,c_2$ and $\bar v_i$ literal in $c_3,c_4$: 
 \medskip
 \begin{center}
\begin{tikzpicture}
\tikzstyle{every node}=[inner sep=0.5mm,draw,circle] 

	\node[draw=none, fill=none] (n) at (1.3,0) {{\small $t_i^1$}};			
	\node (t1) at (1,0) {};
	\node[draw=none, fill=none] (n) at (1.3,-0.5) {{\small $c_2$}};			
	\node (c2) at (1,-0.50) {};
	\node[draw=none, fill=none] (n) at (-0.5,1.3) {{\small $c_3$}};	
	\node (c3) at (-0.5,1) {};		
	\node[draw=none, fill=none] (n) at (0,1.3) {{\small $f_i^1$}};	
	\node (f1) at (0,1) {};	
	\node[draw=none, fill=none] (n) at (2.5,1.25) {{\small $c_4$}};
	\node (c4) at (2.5,1) {};		
	\node[draw=none, fill=none] (n) at (2,1.3) {{\small $f_i^2$}};
	\node (f2) at (2,1) {};	
	\node[draw=none, fill=none] (n) at (0.7,2.5) {{\small $c_1$}};	
	\node (c1) at (1,2.5) {};	
	\node[draw=none, fill=none] (n) at (0.7,2) {{\small $t_i^2$}};	
	\node (t2) at (1,2) {};	
	\node[draw=none, fill=none] (n) at (0.25,0.25) {{\small $ x_i^1$}};	
	\node[fill=black] (x1) at (0.5,0.5) {};	
	\node[draw=none, fill=none] (n) at (1.7,1.7) {{\small $ x_i^2$}};		
	\node[fill=black] (x2) at (1.5,1.5) {};	
				
  \foreach \from/\to in {f1/x1,t1/x1,f2/x2,t2/x2, t1/t2,c3/f1,c2/t1,f2/c4,t2/c1}
    \draw (\from) -- (\to);
\end{tikzpicture}
\end{center} 
All other cases are rotations of the above three cases and/or invert the roles of $v_i$ and $\bar v_i$. Also, if a variable only appears positively (or negatively), it can be deleted along with the clauses which contain it. The maximum degree of the vertices which replace $v_i$ is three.

Replace each clause-vertex $c_j$ by the following subgraph:
 \medskip
 \begin{center}
\begin{tikzpicture}
\tikzstyle{every node}=[inner sep=0.5mm,draw,circle] 
	\node[draw=none, fill=none] (n) at (1,-0.3) {{\small $c_j^1$}};			
	\node (cj1) at (1,0) {};
	\node[draw=none, fill=none] (n) at (1,2.3) {{\small $c_j^2$}};			
	\node (cj2) at (1,2) {};
	\node[draw=none, fill=none] (n) at (0,0.2) {{\small $z_j^1$}};			
	\node (zj1) at (0,0.5) {};
	\node[draw=none, fill=none] (n) at (0,1.8) {{\small $z_j^2$}};			
	\node (zj2) at (0,1.5) {};
	\node[draw=none, fill=none] (n) at (-1,1.25) {{\small $z_j$}};			
	\node[fill=black] (zj) at (-1,1) {};				
	\node[draw=none, fill=none] (n) at (-1.5,1.25) {{\small $s_j$}};			
	\node[fill=black] (s) at (-1.5,1) {}; 
 	\node[draw=none, fill=none] (n) at (-2,1.25) {{\small $p_j$}};			
	\node (p) at (-2,1) {};
 
  \foreach \from/\to in {cj1/zj1,cj2/zj2,zj1/zj,zj2/zj,zj/s,s/p}
    \draw (\from) -- (\to);
\end{tikzpicture}
\end{center} 
The vertices $c_j^1,c_j^2$ somehow take the role of the old vertex $c_j$ regarding its neighbours:  $c_j^1$ is adjacent to two of the literals of $c_j$ and $c_j^2$ is adjacent to the remaining literal. This way, all vertices have degree at most three and the choices of literals to connect to  $c_j^1$ and $c_j^2$ can be made such that  planarity is preserved. 

Let $G'$ be the graph obtained from $G$ by replacing all vertices according to the above rules. The input $G'$ and $S:=\{x_i^1,x_i^2\colon i=1,\dots, n\}\cup\{s_j,z_j\colon j=1,\dots,m\}$ is a ``yes"-instance for  {\sc Minimal Dominating Set Extension} if and only if the formula associated to $G$ is a ``yes"-instance for 4P3C3SAT.

Let $G$ be the graph associated to a satisfiable 4P3C3SAT-formula $c_1\wedge c_2\wedge \dots \wedge c_m$. Consider a satisfying assignment $\phi$  for $c_1\wedge c_2\wedge\dots\wedge c_m$  and the corresponding vertex-set $W:=\{t_i^1,t_i^2\colon \phi(v_i)=1\}\cup\{f_i^1,f_i^2\colon \phi(v_i)=0\}$.  
Let $W'$ be an arbitrary inclusion-minimal subset of $W$ such that  $\{c^1_j,c^2_j\}\cap N_{G'}(W')\not=\emptyset$ for all $j\in \{1,\dots,m\}$; $W$ itself has this domination-property since $\phi$ satisfies the formula $c_1\wedge c_2\wedge\dots\wedge c_m$. By the inclusion-minimality of $W'$, the set $S\cup W'$ is irredundant: Each vertex in $W'$ has at least one of the $c_j^k$ as private neighbour, the vertices $x_i^k$ have  either $t^k_i$ or $f^k_i$ as a private neighbour, $pn(s_j,S\cup W')=\{p_j\}$ and $pn(z_j,S\cup W')=\{z_j^1,z_j^2\}$. The set $S\cup W$ might however not dominate all vertices $c_j^k$.  Adding the set $Y:=\{z_j^k\colon c_j^k\notin  N_{G'}(W)\}$ to $S\cup W$ creates a dominating set. Since for each clause $c_j$  either $c_j^1\in N_{G'}(W')$ or $c_j^2\in N_{G'}(W')$, either $z_j^1$ or $z_j^2$ remains in the private neighbourhood of $z_j$. Other private neighbourhoods are not affected by $Y$. At last, each vertex $z_j^k\in Y$ has the clause-vertex $c_j^k$ as private neighbour, by the definition of $Y$, so overall $S\cup W'\cup Y$ is a minimal dominating set.

If the input $(G',S)$ is a ``yes"-instance for  {\sc Minimal Dominating Set Extension}, the set $S$ can be extended to a set $S'$ which especially dominates all vertices $c_j^k$ and has at least one private neighbour for each $z_j$. The latter condition implies that  $S'\cap \{z_j^k,c_j^k\}=\emptyset$ for $k=1$ or $k=2$ for each $j\in \{1,\dots,m\}$. A vertex $c_j^k$ for which $S'\cap \{z_j^k,c_j^k\}=\emptyset$ has to be dominated by a variable-vertex, which means that $t_i^k\in S'$ ($f_i^k\in S'$) for some variable $v_i$ which appears positively (negatively) in $c_j$.  Minimality of $S'$ requires at least one private neighbour for each $x_i^k$ which, by construction of the variable-gadgets, means that either $\{f_i^1,f_i^2\}\cap S'=\emptyset$ or $\{t_i^1,t_i^2\}\cap S'=\emptyset$, so each variable can only be represented  either positively or negatively in $S'$. Overall, the assignment $\phi$ with $\phi(v_i)=1$ if $\{t_i^1,t_i^2\}\cap S'\not=\emptyset$  and $\phi(v_i)=0$ otherwise satisfies $c_1\wedge c_2\wedge\dots\wedge c_m$.

 Finally, $G'$ can be transformed into a cubic planar graph, by adding the following subgraph once to every vertex $v$ of degree two, and twice for each degree one vertex:
 \begin{center}
\begin{tikzpicture}
\tikzstyle{every node}=[inner sep=0.5mm,draw,circle] 
	\node[draw=none, fill=none] (n) at (0.2,1.5) {{\small $v$}};			
	\node (v) at (0,1.5) {};
	\node (y) at (0,0.5) {};
	\node (w2) at (0,1) {};
	\node[fill=black] (w5) at (0.55,0.5) {};
	\node[fill=black] (w3) at (-0.5,0.5) {};	
	\node(w4) at (0,0) {};

  \foreach \from/\to in {w2/w3,w2/w5,w5/w4,w4/w3,w5/y,w3/y,y/w4,w2/v}
    \draw (\from) -- (\to);
\end{tikzpicture}
\end{center}
Add the new black vertices to the set $S$. Then all new vertices are dominated and adding another one of them to the dominating set violates irredundance. The original vertex is not dominated, adding it to the dominating set does not violate irredundance within the new vertices and the new vertices can never be private neighbours to any original vertex so the structure of $G'$ in the above argument does not change.}

\begin{pf}\proofofTheoremMDSEhardness
\end{pf}

\section{General graphs}

It has been already shown decades ago that \UD is NP-complete on general graphs. Also, polynomial-time solvable graph classes are known, mainly due to the fact that on certain graph classes like bipartite graphs, $\alpha(G)=\Gamma(G)$ holds, and it is known that
on such classes, the independence number can be computed in polynomial time. We refer to the textbook on domination~\cite{HHS98} for further details.
In this section, we complement these results by according results on the parameterised complexity and approximation complexity of \UD and the complement problem.

\subsection{Parameterised complexity}

\begin{theorem}\label{w_hardness}
\UD is W[1]-hard.
\end{theorem}

Our proof is a reduction from 
\MCC, a problem introduced in~\cite{FelHRV2009} to facilitate W[1]-hardness proofs. We leave it open whether our problem
belongs to W[1] or whether it can be shown to 
be W[1]-hard on very restricted graph classes,
similar to the results obtained in~\cite{FelHRV2009}
for \textsc{Minimum Domination}. \newcommand{\proofofTheoremWhardness}{Let $G=(V,E)$  be a graph with $k$ different colour-classes given by $V=V_1\cup V_2\cup \dots\cup V_k$. \MCC asks if there exists a clique $C\subseteq V$ in $G$ such that $|V_i\cap C|= 1$  for all $i=1,\dots,k$. For this problem, one can assume that each set $V_i$ is an independent set in $G$, since edges between vertices of the same colour-class have no impact on the existence of a solution. \MCC is known to be W[1]-complete, parameterised by $k$. We construct a graph $G'$ such that $G'$ has an upper dominating set of cardinality (at least) $k+\frac 12(k^2-k)$ if and only if $G$  is a ``yes"-instance for \MCC which proofs W[1]-hardness for \UD, parameterised by $\Gamma(G')$.

Consider $G'=(V',E')$ given by: $V':=V\cup \{v_e\colon e\in  E\}$ and
\begin{eqnarray*}
E'&:=&\bigcup_{i=1}^kV_i\times V_i\\
&\cup&\bigcup_{i=1}^k\bigcup_{j=1}^k\left\{(v_{(u,w)},x)\colon (u,w)\in (V_i\times V_j)\cap E, x\in \left((V_i\cup V_j)-\{u,w\}\right)\right\}\\
&\cup&\bigcup_{i=1}^k\bigcup_{j=1}^k\left\{(v_{e},v_{e'})\colon e,e'\in (V_i\times V_j)\cap E\right\}\,.
\end{eqnarray*}

If $C\subset V$ is a (multi-coloured) clique of cardinality $k$ in $G$, the set $S':= C\cup \{v_{(u,v)}\colon u,v\in C\}$ is an upper dominating set for $G'$ of cardinality $k+\frac 12(k^2-k)$: First of all, $\{v_{(u,v)}\colon u,v\in C\}\subset V'$ since $(u,v)\in E$ for all $u,v\in C$. Further, by definition of the edges $E',$ $u,v\notin N_{G'}(v_{(u,v)})$ and $u\notin N_{G'}(v)$ for $u$  and $v$ from different colour classes so  $S'$ is an independent set in $G'$ and hence a minimal dominating set. 
It can be easily verified that 
$S'$ 
is also dominating for $G'$ -- observe that it contains exactly one vertex for each clique in the graph.

Suppose $S$ is a minimal dominating set for $G'$. Consider the partition $S=\left(\bigcup _{i=1}^k S_i\right) \cup \left( \bigcup_{1\leq i<j\leq k} S_{\{i,j\}} \right)$ defined by: $S_i:=S\cap V_i$ for $i=1,\dots,k$ and $S_{\{i,j\}}:= S\cap \{v_e\colon e\in V_i\times V_j\}$ for all $1\leq i<j\leq k$. The minimality of $S$ gives the following properties for these subsets of $S$:

\begin{enumerate}
\item If $|S_i|>1$ for some index $i\in \{1,\dots,k\}$, minimality implies $|S_i|=2$ and for all $j\not=i$ either $S_{\{i,j\}}=\emptyset $ or $S_j=\emptyset$:
 \begin{center}
\includegraphics[scale=0.5]{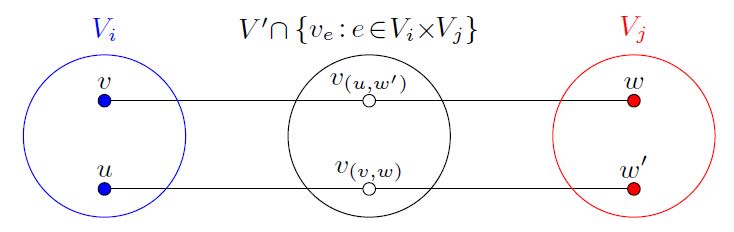}
\end{center}

Since for every $u\in V_i$ and every $j$, $j\not =i$, by construction $V_i\subset N[u]$, and if there is more than one vertex in $S_i$, then their private neighbours have to be in $\{v_e\colon e\in E\}$. A vertex $v_e$ with $e\in V_i\times V_j$ is not adjacent to a vertex $u\in V_i$ if and only if $e=(u,w)$ for some $w\in V_j$. For two different vertices $u,v\in V_i$ consequently all $v_e$ with $e\in V_i\times V_j$ are adjacent to either $u$ or $v$, a third vertex $w\in V_i$ consequently can not have any private neighbour. This also means that any vertex $v_e\in S_{\{i,j\}}$ has to have a private neighbour in $V_j$, so if $S_{\{i,j\}}\not=\emptyset$ the set $S_j$ has to be empty because one vertex from $S_j$ dominates all vertices in $V_j$. These observations hold for all $j\not= i$.

\item If $|S_{\{i,j\}}|>1$ for some indices $i,j\in \{1,\dots,k\}$ we find that $|S_{\{i,j\}}|=2$, $|S_i|,|S_j|\leq 1$ and that $S_i\not=\emptyset$ implies $S_j=S_{\{j,l\}}=\emptyset$ for all $l\in \{1,\dots,k\}-\{i,j\}$ (and equivalently $S_j\not=\emptyset$  implies $S_i=S_{\{i,l\}}=\emptyset$ for all $l\in \{1,\dots,k\}-\{i,j\}$):

 \begin{center}
\includegraphics[scale=0.4]{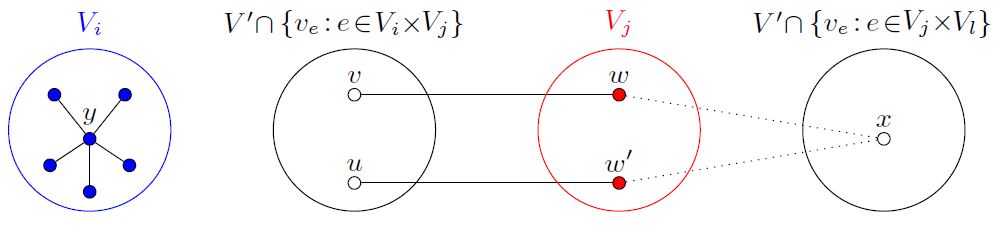}
\end{center}

 Since for any two vertices $u,v$ from $S_{\{i,j\}}$ we have $\{v_e\colon e\in (V_i\times V_j)\cap E\}\cup V_i\cup V_j \subset N(u)\cup N(v)$, the cardinality of $S_{\{i,j\}}$ can be at most two. If there is a vertex $y$ in $S_i$, it already dominates all of $V_i$ so private neighbours for $u,v\in S_{\{i,j\}}$ have to be in $S_j$. For any two vertices $w,w'\in V_j$  any $v_e\in V'\cap\{v_e\colon e\in V_j\times V_l, \ l=1,\dots,k\}$ is either adjacent to at least $w$ or $w'$, so especially for the private vertices of $u$ and $v$ every $x\in S_{j,l}$ would be adjacent to one of them and can consequently not be in a minimal dominating set, so $S_j=S_{\{j,l\}}=\emptyset$. Dominating the vertices in $S_{\{j,l\}}$ for $l\not=i$ then requires $|S_l|=2$ for all $l\not=i$, which leaves no possible private vertices outside $V_i$ for vertices in $V_i$, so $|S_i|=1$.

\item  If $|S_i|=2$  there exists an index $j\not=i$ such that $S_{\{i,j\}}=\emptyset$ and $|S_j|\leq 1$:\\
Let $u,v\in S_i$.  
By the structure of $G'$,  $u$ and $v$ share all neighbours in $V_i$ and $v_e$ such that $e=(x,y)\in V_i\times V_l$ with $x\not\in \{u,v\}$ for all $l\not=i$, so especially the private neighbourhood of $u$ is restricted to $pn(u,S)\subseteq\{v_e\colon e=(v,y)\in E\}$. 
 Let $j$ be an index such that there is a vertex $z\in V_j$ with $v_{(u,z)}\in pn(v,S)$ (there is at least one such index). No neighbours of $v_{(u,z)}$ beside $v$ can be in $S$, which means that $S_{\{i,j\}}=\emptyset$ and $S_j\subseteq \{z\}$.

\item $|S_{\{i,l\}}|= 2$ implies $|S_{\{j,l\}}|\leq 1$ for all $j\not=i$.\\
Suppose $|S_{\{i,l\}}|,|S_{\{j,l\}}|\geq 2$ for some  indices $i,j,l\in \{1,\dots,k\}$.
By property 2 both sets $S_{\{i,l\}},S_{\{j,l\}}$ have cardinality two so let $u_i,w_i\in S_{\{i,l\}} $ and $u_j,w_j\in S_{\{j,l\}}$. Since  
each set $\{v_e\colon e\in E\cap (V_s\times V_t)\}$ is a clique, the private neighbours for these vertices have to be in $V_i,V_j,V_l$. Suppose $v\in pn(u_i,S)\cap V_l$ which means that $w_i,u_j,w_j$ are not adjacent to $v$. This is only possible if $w_i$ represents some edge $(v,x)\in E\cap V_l\times V_i$ and $u_j,w_j$ represent some edges $(v,y),(v,y')\in E\cap V_l\times V_j$. By definition of $E'$, $w_i,u_j,w_j$ 
then share their neighbourhood in $V_l$ (namely $V_l-\{v\}$) which means that $pn(w_i,S)\subset V_i$ 
and $pn(u_j)\cup pn(w_j)\subset V_j$ which implies $S_i=S_j=\emptyset$. So in any case, even if there is no $v\in pn(u_i,S)\cap V_l$, at least one of the sets $V_i$ or $V_j$ contains two vertices which are private neighbours 
for $S_{\{i,j\}}$ and $S_i=S_j=\emptyset$.

Suppose $V_j$ contains two private vertices $y\not=y'$ for $u_j$ and $w_j$ respectively. For any two arbitrary vertices $n_1,n_2\in V_j$, any vertex $x\in\{v_e\colon e\in E\cap (V_i\times V_j)\}$ is adjacent to at least one of them, which means that any $x\in S_{\{i,j\}}$ would steal at least $y\in pn(u_j)$ or $y'\in pn(w_j)$ as private neighbour. 
Minimality of $S$ hence demands $S_i=S_j=S_{\{i,j\}}=\emptyset$. A set with this property however does not dominate any of the vertices $v_e$ with $e\in  E\cap (V_i\times V_j)$. (The set $E\cap (V_i\times V_j)$ is not empty unless the graph $G$ is a trivial ``no"-instance for \MCC.)

\end{enumerate}
According to these properties, the indices of these subsets of $S$ can be divided into the following six sets: $C_i:=\{j\colon |S_j|=i\}$ and $D_i:=\{(j,l)\colon |S_{\{j,l\}}|=i\}$ for $i=0,1,2$ which then give $|S|=2(|C_2|+|D_2|)+|C_1|+|D_1|$. If $|C_2|+|D_2|\not=0$ and $k>3$, we can construct an injective mapping $f\colon C_2\cup D_2 \cup \{a\}\rightarrow C_0\cup D_0$ 
with some $a\notin V'$ in the following way:
\begin{itemize}
\item For every $i\in C_2$ choose some $j\not=i$ with $(i,j)\in D_0$ and $j\notin C_2$   which exists according to property 3 and set $f(i)=(i,j)$. Since  $j\notin C_2$  this setting is injective.

If $D_2=\emptyset$ and $C_2=\{i\}$, choose some $l\not=i$ and map $a$ via $f$ either to $l$ or to $(i,l)$, since, by property 1, one of them is in $C_0$ or $D_0$ respectively. If $D_2=\emptyset$ and $|C_2|>1$, choose some $i,l\in C_2$ and set $f(a)=(i,l)$ since $S_{\{i,l\}}=\emptyset$ by property 1 and neither $i$ nor $l$ is mapped to $(i,l)$.
 
\item For  $(i,j)\in D_2$, property 2 implies at least $i$ or $j$ lies in $C_0$. By property 4  we can choose one of them arbitrarily without violating injectivity. If both are in $C_0$ we can use one of them to map $a$. If for all $(i,j)\in D_2$ only one of the indices $i,j$ is in $C_0$, we still have to map $a$, unless $f(a)$ has been already defined. Assume for $(i,j)\in D_2$ that $i\notin C_0$. By property 2 $\{(j,l)\colon l\notin\{i,j\}\}\subset D_0$. If we cannot choose one of these index-pairs as injective image for $a$, they have all been used to map $C_2$ which means $\{1,\dots,k\}-\{i,j\}\subseteq C_2$ and hence, by property 1, all index-pairs $(l,h)$ with $l,h\in\{1,\dots,k\}-\{i,j\}$ are in $D_0$ and so far not in the image of $f$, so we are free to chose one of them as image of $a$, unless $f(a)$ has been already defined.
\end{itemize}
This injection proves that $|C_2|+|D_2|>0$  implies that $|C_2|+|D_2|<|C_0|+|D_0|$. This means that, regardless of the structure of the original graph $G$, the subsets $S_i$ and $S_{i,j}$ of $S$ either all contain exactly one vertex or 
 $k+\frac 12 (k^2-k)=|C_1|+|D_1|+|C_0|+|D_0|+|C_2|+|D_2|>|C_1|+|D_1|+2(|C_2|+|D_2|)=|S|$. 
 
 So if $|S|= k+\frac 12(k^2-k)$, the above  partition into the sets $S_i,S_{i,j}$ satisfies $|S_i|=|S_{\{i,j\}}|=1$ for all $i,j$. A set with this property is always dominating for $G'$ but only minimal if each vertex has a private neighbour. For some  $v_e\in S_{\{i,j\}}$ this implies that there is some private neighbour $e'=(u,v)\in V'\cap(V_i\times V_j)$ that is not dominated by the (existing) vertex $u'$ in $S_i$ or the vertex $v'$ in $S_j$; (all vertices $V_i$ and $V_j$ are already dominated by $\{u',v'\}\subset S$ and cannot be private neighbours for $v_e$). By construction of $E'$, this is only possible if $(u,v)=(u',v')\in E$. Since this is true for all  index-pairs $(i,j)$, the vertices $\{v\colon v\in S_i, i=1,\dots,k\}$ form a clique in the original graph $G$.}
\begin{pf}\proofofTheoremMDSEhardness
\end{pf}
We do not know if \UD belongs to W[1], but we can at least place it in W[2], the next level of the W hierarchy.\SV{ We obtain this result by describing a suitable multi-tape Turing machine that solves this problem.}

\begin{proposition}\label{prop-Wtwo}
\UD belongs to W[2].
\end{proposition}

\newcommand{\proofofPropWtwo}{(Sketch)
Recall how \MD can be seen to belong to W[2] by providing
an appropriate multi-tape Turing machine\LV{~\cite{Ces2003}}\SV{\footnote{Confer  M.~Cesati. The {T}uring way to parameterised complexity. {\em Journal of Computer and System Sciences}, 67:654--685, 2003.}\ }. 
First,
the $k$ vertices that should belong to the dominating set are guessed, and then this guess is verified in $k$ further (deterministic) steps using $n$ further tapes in parallel,
where $n$ is the order of the input graph. 
We only need to make sure that the guessed set of vertices is minimal.
To this end, we copy the guessed vertices $k$ times, leaving one out each time, and we also guess one vertex for each of the $k-1$-element sets that is not dominated by this set.
Such a guess can be tested in the same way as sketched before using 
parallel access to the $n+1$ tapes.
The whole computation takes $O(k^2)$ parallel steps of the Turing machine, which shows membership in W[2].}
\begin{pf}\proofofPropWtwo
\end{pf}

Another interesting question  is to consider the dual parameter $\ell$, that is to decide the existence of an upper  dominating set  of size at least $n-\ell$. This is in fact the natural parameterisation for \CUD.
\SV{By some combinatorial arguments exploiting some $(F,I,P.O)$ decomposition implied by an upper dominating set, we can prove:}

\begin{theorem}\label{cud_kernel}
\CUD is in FPT. More precisely, it admits a kernel of at most  $\ell^2+\ell$ many vertices and at most $\ell^2$ many edges.
\end{theorem}
\newcommand{\proofofTheoremcudkernel}{Let $G=(V,E)$ be an arbitrary input graph with $|V|=n$.
First consider a vertex $v\in V$  with $deg(v)>\ell$ and any minimal dominating set $D$ with some associated partition $(F,I,P,O)$:
\begin{itemize}
 \item If $v\in I$, all  neighbours of $v$ have to be in $O$ which means $|O|\geq|N(v)|>\ell$.
 \item If $v\in F$, exactly one neighbour $p$ of $v$ is in $P$ and $N[v]-\{p\}\subseteq F\cup O$, which gives $|O|+|P|=|O|+|F|\geq |N[v]-\{p\}|>\ell$.
 \item If $v\in P$, exactly one neighbour $p$ of $v$ is in $F$ and $N[v]-\{p\}\subseteq P\cup O$, so $|O|+|P|>\ell$.
 \end{itemize}
We always have either $v\in O$ or $|O|+|P|>\ell$, which means a ``no"-instance for \CUD. Consider the graph $G'$ built from $G$ by deleting the vertex $v$ and all its edges. For any minimal dominating set $D$ for $G$ with partition $(F,I,P,O)$ such that $v\in O$, $D$ is also minimal for $G'$, since $pn(w,D)\supseteq\{w\}$ for all $w\in I$ and $|pn(u,D)\cap P|=1$ for all $u\in F$. Also, any set $D'\subset V-\{v\}$ which does not dominate $v$ has a cardinality of at most $|V-N[v]|<n-\ell$, so if $G'$ has a dominating set $D'$ of cardinality at least $n-\ell$, $N(v)\cap D'\not= \emptyset$, so $D'$ is also dominating for $G$. These observations allow us to successively reduce $(G,\ell)$ to $(G',\ell')$ with $\ell'=\ell -1$ as long as there are vertices $v$ with $deg(v)>\ell$, similar to Buss's rule for parameterised \MVC. Any isolated vertex in the resulting graph $G'$ originally only has neighbours in $O$ which means it belongs to $I$ in any dominating set $D$  with partition $(F,I,P,O)$ and can hence be deleted from $G'$ without affecting the existence of an upper dominating set with $|P|+|O|\leq \ell'$. 

Let $(G',\ell')$ be the instance obtained after the reduction for all vertices of degree larger than $\ell$ and after deleting isolated vertices with $G'=(V',E')$ and let $n'=|V'|$. If there is an upper dominating set $D$ for $G'$ with $|D|\geq n'-\ell'$, any associated partition $(F,I,P,O)$ for $D$ satisfies $|P|+|O|\leq \ell'$. Since $G'$ does not contain isolated vertices, every vertex in $I$ has at least one neighbour in $O$. Also, any vertex in $V'$, and hence especially any vertex in $O$, has degree at most $\ell'$, which means that $|I|\leq |N(O)|\leq \ell'|O|$. Overall:
\begin{equation}\label{eq-cud_kernel}
|V'|\leq |I|+|F|+|P|+|O|\leq (\ell'+1)|O|+2|P|\leq \max_{j=0}^{\ell'}\{j(\ell'+1),2(\ell'-j)\}\,,
\end{equation}
and hence $|V'|\leq \ell'(\ell'+1)$, 
or $(G',\ell')$ and consequently $(G,\ell)$ is a ``no"-instance.
Concerning the number of edges, we can derive a similar estimate.
There are at most $\ell$ edges incident with each vertex in $O$.
In addition, there is exactly one edge incident with each vertex in $P$ that has not yet been accounted for, and, in addition, there could be $\ell-1$ edges incident to each vertex in $F$ that have not yet been counted. This shows the claim.
}
\begin{pf} \proofofTheoremcudkernel
\end{pf}

\smallskip

\LV{We just derived a kernel result for \CUD, in fact a kernel of quadratic size in terms of the number of vertices and edges.} 
This \SV{quadratic-size kernel}
poses the natural question if we can do better.
Next, the question is if the brute-force search we could perform on the quadratic kernel is the best we can do to solve \CUD in FPT time. Fortunately, this is not the case, as the following result shows.

\SV{\begin{proposition}\label{prop-CoUDbranching}
\textsc{Co-Upper Domination} can be solved in time $O^*(4.3077^\ell)$.
\end{proposition}

\begin{prf} (Sketch)
This result can be shown by designing a branching algorithm that takes a graph $G=(V,E)$ and a parameter $\ell$ as input.
%
As in Section~\ref{sec-FIPO}, 
to each graph $G=(V,E)$ and (partial) dominating set, we associate a partition ($F,I,P,O$).
We consider $\kappa = \ell - (\frac{|F|}{2} + \frac{|P|}{2} + |O|)$ as a measure
of the partition. Note that $\kappa \leq \ell$.
At each branching step, our algorithm picks some vertices from $R$ (the set of yet undecided vertices).
They are either added to the current dominating set $D := F \cup I$ or to $\overline{D} := P \cup O$.
Each time a vertex is added to $P$ (resp. to $O$) the value of $\kappa$ decreases by $\frac{1}{2}$ (resp. by $1$).
Also, whenever a vertex $x$ is added to $F$, the value of $\kappa$ decreases by $\frac{1}{2}$.

Let us describe the two halting rules. First, whenever $\kappa$ reaches zero, we are face a ``no''-instance.
Then, if the set $R$ of undecided vertices is empty, we check whether the current domination set $D$
is minimal and of size at least $n- \ell$, and if so, the instance is a ``yes''-instance.
Then, we have a simple reduction rule: whenever the neighbourhood of a undecided vertex $v \in R$ is included
in $\overline{D}$, we can safely add $v$ to $I$.
Finally, vertices are placed to $F$, $I$ or $\overline{D}$ according to three branching rules.
The first one considers undecided vertices with a neighbour already in $F$ (in such a case, $v$ cannot belongs to $I$).
The second one considers undecided vertices with only one undecided neighbour (in such a case, several
cases may be discarded as, e.g., they cannot be both in $I$ or both in $\overline{D}$).
The third branching rule considers all the  possibilities for an undecided vertex and due to
the previous branching rules, it can be assumed that each undecided vertex has at least two
undecided neighbours (which is nice since such vertices have to belong to $\overline{D}$ whenever
an undecided neighbour is added to $I$).
\end{prf}}

\newcommand{\CoUDbranching}{
\begin{proposition}
Given a graph $G=(V,E)$ and a parameter $\ell$, a call of Algorithm \texttt{ComputeCoUD} with
parameters ($G$, $\ell$, $\emptyset$, $\emptyset$, $\emptyset$, $\ell$)
solves \textsc{Co-Upper Domination} in time $O^*(4.3077^\ell)$.
\end{proposition}

\begin{algorithm}[h]
	\caption{\label{algo-StableMax}\texttt{ComputeCoUD($G$, $\ell$, $F$, $I$, $\overline{D}$, $\kappa$)}}
	\DontPrintSemicolon
	\SetKwInOut{Input}{input}\SetKwInOut{Output}{output}

	{\footnotesize
	\Input{\textit{a graph $G=(V,E)$, parameter $\ell \in \mathbb{N}$, three disjoint sets $F,I,\overline{D} \subseteq V$ and $\kappa \leq \ell$.}}
	\Output{\textit{``yes'' if $\Gamma(G) \geq |V| - \ell$; ``no'' otherwise.}}
	}
	\smallskip
	Let $R \gets V \setminus (F \cup I \cup \overline{D})$\;
	\lIf(\hfill {(H1)}){$\kappa < 0$}
	{
		\Return ``no''
	}	
	\If(\hfill {(H2)}){$R$ is empty}
	{
		\If{$F \cup I$ is a minimal dominating set of $G$ and $|F \cup I| \geq n-\ell$}
		{
			\Return ``yes''\;
		}
		\lElse
		{
			\Return ``no''\;
		}
	}
	\If(\hfill {(R1)}){there is a vertex $v \in R$ s.t. $N(v) \subseteq \overline{D}$}
	{
		\Return \texttt{ComputeCoUD($G$, $\ell$, $F$, $I \cup \{v\}$, $\overline{D}$, $\kappa$)}\;
	}

	\If(\hfill {(B1)}){there is a vertex $v \in R$ s.t. $|N(v) \cap F| \geq 1$}
	{
		\Return \texttt{ComputeCoUD($G$, $\ell$, $F \cup \{v\}$, $I$, $\overline{D}$, $\kappa-\frac{1}{2}$)} $\lor$\linebreak[2] \texttt{ComputeCoUD($G$, $\ell$, $F$, $I$, $\overline{D} \cup \{v\}$, $\kappa-\frac{1}{2}$)}\;
	}
	
	\If(\hfill {(B2)}){there is a vertex $v \in R$ s.t. $|N(v) \cap R| = 1$}
	{
		Let $u$ be the unique neighbour of $v$ in $R$\;
		\Return \texttt{ComputeCoUD($G$, $\ell$, $F \cup \{u,v\}$, $I$, $\overline{D}$, $\kappa-1$)} $\lor$ \texttt{ComputeCoUD($G$, $\ell$, $F \cup \{u\}$, $I$, $\overline{D} \cup \{v\}$, $\kappa-1$)} $\lor$\linebreak[2] \texttt{ComputeCoUD($G$, $\ell$, $F$, $I \cup \{v\}$, $\overline{D} \cup \{u\}$, $\kappa-1$)}\;
	}

	\Else(\hfill {(B3)})
	{
		Let $v$ be a vertex of $R$\;
		\Return \texttt{ComputeCoUD($G$, $\ell$, $F$, $I \cup \{v\}$, $\overline{D} \cup N(v)$, $\kappa-2$)} $\lor$\linebreak[2] \texttt{ComputeCoUD($G$, $\ell$, $F \cup \{v\}$, $I$, $\overline{D}$, $\kappa-\frac{1}{2}$)} $\lor$\linebreak[2] \texttt{ComputeCoUD($G$, $\ell$, $F$, $I$, $\overline{D} \cup \{v\}$, $\kappa-\frac{1}{2}$)}\;
	}
\end{algorithm}

\begin{prf}
Algorithm \texttt{ComputeCoUD} is a branching algorithm, with halting rules (H1) and (H2), reduction
rule (R1), and three branching rules (B1)-(B3).
We denote by $G=(V,E)$ the input graph and by $\ell$ the parameter. At each call, the set of vertices $V$
is partitioned into four sets: $F$, $I$, $\overline{D}$ and $R$. The set of ``remaining'' vertices
$R$ is equal to $V \setminus (F \cup I \cup \overline{D})$,
and thus can be obtained from $G$ and the three former sets.

At each recursive call, the algorithm picks some vertices from $R$. They are either added to the current dominating set $D := F \cup I$,
or to the set $\overline{D}$ to indicate that they do not belong to any extension of the current dominating set.
The sets $F$ and $I$ are as previously described (\emph{i.e.}\, if we denote by $D$ the dominating set we are looking for,
$I:=\{v\in D\colon v\in pn(v,D)\}$ and $F:=D-I$).

\medskip

Note that parameter $\kappa$ corresponds to our ``budget'', which is initially set to $\kappa := \ell$.
Recall that any minimal dominating set of a graph $G=(V,E)$ can be associated with a partition ($F,I,P,O$)
(see Section~\ref{sec-FIPO} for the definitions of the sets and for some properties).
If we denote by $D$ a minimal dominating set of $G$ and by $\overline{D}$ the set $V \setminus D$,
then by definition, $F, I$ is a partition of $D$ and $P,O$ is a partition of $\overline{D}$.
Also, by definition of $F$ and $P$, it holds that $|F|=|P|$ and there is a perfect matching between vertices of $F$ and $P$.
Since each vertex of $F$ will (finally) be matched with its private neighbour from $P$,
we define our budget as $\kappa = \ell - (\frac{|F|}{2} + \frac{|P|}{2} + |O|)$.
One can observe that if $D$ is a minimal dominating set of size at least $n-\ell$ then $\kappa \geq 0$.
Conversely, if $\kappa < 0$ then any dominating set $D$ such that $F \cup I \subseteq D$
is of size smaller than $n - \ell$. This shows the correctness of \textbf{(H1)}.
We now consider the remaining rules of the algorithm.
Note that by the choice of $\kappa$, each time a vertex $x$ is added to $\overline{D}$,
the value of $\kappa$ decrease by $\frac{1}{2}$ (or by $1$ if we can argue that $x$ is not matched with a vertex of $F$
and thus belongs to $O$). Also, whenever a vertex $x$ is added to $F$, the value of $\kappa$ decrease by $\frac{1}{2}$.

\begin{description}
\item[(H2)] If $R$ is empty, then all vertices have been decided: they are either in $D:=F\cup I$ or in $\overline{D}$.
It remains to check whether $D$ is a minimal dominating set of size at least $n-\ell$.
\item[(R1)] All neighbours (if any) of $v$ are in $\overline{D}$ and thus $v$ has to be in $I \cup F$.
As $v$ will also belong to $pn(v,D)$, we can safely add $v$ to $I$.
Observe also that this reduction rule does not increase our budget.
\item[(B1)] Observe that if $v$ has a neighbour in $F$, then $v$ cannot belong to $I$.
When a vertex $v$ in added to $F$ the budget is reduced by at least $\frac{1}{2}$;
when $v$ is added to $\overline{D}$, the budget is reduced by $\frac{1}{2}$ as well.
So (B1) gives a branching vector of $(\frac{1}{2}, \frac{1}{2})$.
\item[(B2)] If (R1) and (B1) do not apply and $N(v)\cap R=\{u\}$, then the vertex $v$ has to either
dominate itself or be dominated by $u$. Every vertex in $F$ has a neighbour in $F$, which in this
case means that $v\in F$ implies $u\in F$ (first branch). Moreover, the budget is the reduced
by at least $2 \cdot \frac{1}{2}$.

If $v$ is put in $I$, $u$ has to go to $\bar D$ (third branch). Thus $u$ cannot be a private neighbour of some $F$-vertex,
and the budget decreases by at least $1$ ($u \in O$).

If $v$ does not dominate itself, $u$ has to be in $F\cup I$. In this last case
it suffices to consider the less restrictive case $u\in F$, as $v$ can be chosen as the private neighbour
for $u$ (second branch). If $u$ is indeed in $I$ for a minimal dominating set which extends the current $I\cup F$,
there is a branch which puts all the remaining neighbours of $u$ in $\bar D$. Observe that we only
dismiss branches with halting rule (H2) where we check if $F\cup I$ is a minimal dominating set,
we do not require the chosen partition to be correct. As for the counting in halting rule (H1):
weather we count $u\in F$ and $v\in P$ (recall that $P \subseteq \overline{D}$) each with $\frac{1}{2}$
or count $v\in O$ (recall that $O \subseteq \overline{D}$) with $1$ does not make a difference for $\kappa$.
So the budget decreases by at least $1$.

Altogether (B2) gives a branching vector of $(1, 1, 1)$.

\item[B3] The correctness of (B3) is easy as all possibilities are explored for vertex $v$.
Observe that by (R1) and (B2), vertex $v$ has at least two neighbours in $R$.
When $v$ is added to $I$, these two vertices are removed (and cannot be the private neighbours of some $F$-vertices).
Thus we reduce the budget by at least $2$.
When $v$ is added to $F$, the budget decreases by at least $\frac{1}{2}$.
When $v$ is added to $\overline{D}$, we reduce the budget by at least $\frac{1}{2}$.
Thus (B3) gives a branching vector of $(2, \frac{1}{2}, \frac{1}{2})$. However, we can observe
that the second branching rule (\emph{i.e.,} when $v$ is added to $F$) implies
a subsequent application of (B1) (or rule (H1) would stops the recursion). Thus the branching
vector can be refined to $(2, 1, 1, \frac{1}{2})$.
\end{description}

Taking the worst-case over all branching vectors, establishes the claimed running-time.
\end{prf}
}
\LV{\CoUDbranching{}}

Of course, the question remains to what extent the previously presented parameterised algorithm can be improved on. 

\LV{\todo[inline]{HF: We might want to look into ETH-based lower bounds for the parameterised or exact algorithms.}}

\newcommand{\MMHS}{\textsc{Maximum Minimal Hitting Set}\xspace}

\subsection{Approximation}

Also in this case, the two problems \UD and \CUD behave quite differently, the first one is hard, while the second one is relatively easy to approximate. 

Let us first show that \UD is hard to approximate. We will establish this in
two steps: first, we show that a related natural problem, \MMHS, is hard to
approximate, and then we show that this problem is essentially equivalent to
\UD. 

The \MMHS problem is the following one: we are given a hypergraph, that is, a base
set $V$ and a collection $F$ of subsets of $V$. We wish to find a set
$H\subseteq V$ such that:
\begin{enumerate}
\item For all $e\in F$ we have $e\cap H\neq \emptyset$ (i.e., $H$ is a hitting set).
\item For all $v\in H$ there exists $e\in F$ such that $e\cap H=\{v\}$ (i.e., $H$ is
minimal).
\item $H$ is as large as possible.
\end{enumerate}
It is not hard to see that this problem generalises \UD: given a graph
$G=(V,E)$, we can produce a hypergraph by keeping the same set of vertices and
creating a hyperedge for each closed neighbourhood  $N[v]$ of $G$. An upper
dominating set of the original graph is now exactly a minimal hitting set of
the constructed hypergraph. We will also show that \MMHS can be reduced to \UD.

Let us note that \MMHS, as defined here, also generalises {\sc Maximum Minimal
Vertex Cover}\SV{, which corresponds to instances where the input hypergraph is
actually a graph}. We recall that for this problem there exists a
$n^{1/2}$-approximation algorithm, while it is known to be
$n^{1/2-\varepsilon}$-inapproximable \cite{BorCroPas2013}. Here, we generalise
this result to arbitrary hypergraphs, taking into account the sizes of the
hyperedges allowed.
\SV{The proof of the next theorem is a reduction from \MIS,
compare~\cite{Has99}.}

\begin{theorem}\label{mmhs}

For all $\varepsilon>0, d\ge 2$, if there exists a polynomial-time
approximation algorithm for \MMHS which on hypergraphs 
$G=(V,E)$ 
where hyperedges have
size at most $d$ has approximation ratio $n^{\frac{d-1}{d}-\varepsilon}$, where $|V|=n$, then
P=NP. This \SV{is still true for}\LV{statement still holds if we restrict the problem to} hypergraphs
where $|F|=O(|V|)$.
\end{theorem}

\newcommand{\proofofTheoremmmhs}{Fix some constant hyperedge size $d$. We will present a reduction from \MIS,
which is known to be, in a sense,  completely inapproximable~\cite{Has99}. In particular, it
is known that, for all $\varepsilon>0$, it is NP-hard to distinguish for an
$n$-vertex graph $G$ if $\alpha(G)>n^{1-\varepsilon}$ or
$\alpha(G)<n^{\varepsilon}$.

Take an instance $G=(V,E)$ of \MIS. We begin by considering this graph as a
hypergraph (with size-two hyperedges), to which we will add some more vertices
and hyperedges.  For every set $S\subseteq V$ such that $|S|=d-1$ which is
an independent set in $G$, we add to our instance $n$ new vertices, call them
$u_{S,i}$, $1\le i\le n$. Also, for each such vertex $u_{S,i}$ we add to our
instance the hyperedges $S\cup\{u_{S,i}\}$, $1\le i\le n$. This completes the construction. It
is not hard to see that the constructed hypergraph has hyperedges of size at
most $d$, and its vertex and hyperedge set are both of size $O(n^d)$.

Let us analyse the approximability gap of this reduction. First, suppose that
in the original graph we have $\alpha(G) > n^{1-\varepsilon}$. Then, there
exists a minimal hitting set of the new instance with size at least
$n^{d-O(\varepsilon)}$. To see this, consider a maximum independent set $I$ of
$G$. We set $H$ to be $(V\setminus I) \cup \{ u_{S,i}\ |\ S\subseteq I, 1\le
i\le n \}$. In words, we add to $H$ a minimum vertex cover of $G$, as well as
the $u_{S,i}$ vertices whose neighbourhoods are contained in $I$. It is not hard
to see that $H$ is a hitting set, because each of the size-$d$ hyperedges not
hit by $V\setminus I$ is hit by a unique selected vertex $u_{S,i}$. Because of
this, and since $V\setminus I$ is a minimal vertex cover of $G$, $H$ is also
minimal. Finally, the size of $H$ follows from the fact that there exists
$\alpha(G) \choose d-1$ sets $S$ for which $H$ contains all the vertices
$u_{S,i}$.

For the converse direction, we want to show that if $\alpha(G)<n^\varepsilon$
then any minimal hitting set of the new instance has size at most
$n^{1+O(\varepsilon)}$. Consider a hitting set $H$ of the new instance.  It
must be the case that $H\cap V$ is a vertex cover of $G$, and therefore
$V\setminus H$ is an independent set of $G$. Let $S\subset V$ be a set of
vertices such that $S\cap H\neq \emptyset$. Then $u_{S,i}\not\in H$ for all
$i$, because the (unique) hyperedge that contains $u_{S,i}$ also contains some
other vertex of $H$, contradicting minimality. It follows that $H$ can only
contain $u_{S,i}$ if $S\subseteq V\setminus H$. Because $V\setminus H$ is an
independent set, it has size at most $n^\varepsilon$, meaning there are at most
$n^{\varepsilon}\choose d-1$ sets $S$ such that $H$ may contain vertices
$u_{S,i}$. Thus, the total size of $H$ cannot be more than
$n^{1+O(\varepsilon)}$.}

\begin{pf}\proofofTheoremmmhs
\end{pf}

\begin{corollary}\label{mmhs_apx} 
For any $\varepsilon>0$ \MMHS is not $n^{1-\varepsilon}$-approximable, where $n$
is the number of vertices of the input hypergraph,
unless
P=NP.  
This \SV{is still true for}\LV{statement still holds if we restrict the problem to} hypergraphs
where $|F|=O(|V|)$.
\end{corollary}

\newcommand{\proofofCormmhsapx}{Assume there were, for some  $\varepsilon>0$,
a factor-$n^{1-\varepsilon}$ approximation algorithm $A$ for \MMHS. Then, choose $d$ such that $1/d\leq \varepsilon/2$ and hence $(d-1)/d\geq 1-\varepsilon/2$.
Then, $A$ would be a factor-$n^{(d-1)/d-\varepsilon/2}$
approximation algorithm when restricted to hypergraphs with hyperedges of size at most $d$, contradicting Theorem~\ref{mmhs}.}
\begin{pf}\proofofCormmhsapx
\end{pf}

\begin{theorem}\label{ud_apx} 
For any $\varepsilon>0$ \UD is not $n^{1-\varepsilon}$-approximable,
where $n$ is the number of vertices of the input graph, 
unless
P=NP.  
\end{theorem}

\SV{\begin{prf} (Sketch) 
We construct an $E$-reduction from \MMHS. Given a hypergraph $G=(V,F)$ as an instance
of \MMHS, we define a graph $G'=(V',E')$ as an instance of \UD as follows: $V'$
contains a vertex $v_i$ associated to any vertex $i$ from $V$, a vertex $u_{e}$ for
any edge $e\in F$ and a new vertex $v$. $E'$ contains edges such that $G'[V]$
and $G'[E]$ are cliques. Moreover, $v$ is adjacent to every vertex $v_i \in V$,
and $(v_i,u_{e})\in E'$ if and only if $i\in e$ in $G$.
\end{prf}}

\newcommand{\proofofTheoremudapx}{We construct an $E$-reduction from \MMHS. Given a hypergraph $G=(V,F)$ as an  instance
of \MMHS, we define a graph $G'=(V',E')$ as an instance of \UD as follows: $V'$
contains a vertex $v_i$ associated to any vertex $i$ from $V$, a vertex $u_{e}$ for
any edge $e\in F$  and a new vertex $v$. $E'$ contains edges  such that $G'[V]$
and $G'[E]$ are cliques. Moreover, $v$ is adjacent to every vertex $v_i \in V$,
and $(v_i,u_{e})\in E'$ if and only if $i\in e$ in $G$.

First we show that given a solution $S$ that is a minimal hitting set in $G$,
$S$ is also a minimal dominating set in $G'$. Indeed if $S$ is a hitting set in
$G$ then $S$ is a dominating set in $G'$. If $S$ is minimal, that is, any
proper subset $S'\subset S$ is no longer a hitting set, then it is also the
case that $S'$ is no more a dominating set in $G'$. That implies that $\opt(G')
\geq \opt(G)$.

Consider now an upper dominating set $S$ for $G'$. To dominate the vertex $v$,
$S$ has to contain at least one vertex $w\in V\cup \{v\}$. If $S$ contains one
vertex $u_{e}\in E$, then the set $\{w,u_{e}\}$ is already dominating. If we
want a solution of cardinality more than two, then $S\subseteq V$. If
$S\subseteq V$ is a minimal dominating set in $G'$,  $S$ is also a minimal
hitting set in $G$ since $S$ covers all hyperedges in $G$ if and only if it
dominates all edge-vertices in $G'$. So starting with any minimal dominating
set $S$ of $G'$ of cardinality larger than two, $S$ is also a minimal hitting
set of $G$. 

The result now follows from Corollary \ref{mmhs_apx}.
}
\begin{pf}\proofofTheoremudapx
\end{pf}

\LV{\todo[inline]{HF: We can talk about this; at this moment, I think that this problem fits pretty well into the whole paper, and we do have quite some results for \MMHS: for instance, the non-extensibility immediately transfers from minimal dom set extension,
and also the W[1]-hardness. We might want to mention all this in one sentence.}}

\LV{Let us n}\SV{N}ote that, interestingly, the inapproximability bound given in Theorem
\ref{mmhs} is tight, for every fixed $d$. To see this, consider the algorithm
of the following theorem, which also generalises results
 on {\sc Maximum Minimal Vertex Cover}~\cite{BorCroPas2013}. 
 \LV{To simplify
presentation, we assume that we are given an $n$-vertex hypergraph where every
vertex appears in at least one hyperedge.}

\begin{theorem}\label{mmhs_alg}

For all $d\ge 1$, there exists a polynomial-time algorithm which, given a
hypergraph $G=(V,F)$ such that all hyperedges have size at most $d$, produces a
minimal hitting set $H$ of $G$ with size $\Omega(n^{1/d})$. This shows an  $\Omega(n^{\frac{d-1}{d}})$-approximation for \MMHS on such
hypergraphs.

\end{theorem}

\newcommand{\proofofTheoremmmhsalg}{
The proof is by induction on $d$. For $d=1$, if every vertex appears in a
hyperedge, any hitting set must contain all vertices, so we are done. For
$d>1$, we do the following: first, greedily construct a maximal set $M\subseteq
F$ of pair-wise disjoint hyperedges. If $|M|\geq n^{1/d}$ then we know that any
hitting set of $G$ must contain at least $n^{1/d}$ vertices. So, we simply
produce an arbitrary feasible solution by starting with $V$ and deleting
redundant vertices until our hitting set becomes minimal.

Suppose then that $|M|<n^{1/d}$. Let $H$ be the set of all vertices contained
in $M$, so $|H|<d|M| = O(n^{1/d})$. Clearly, $H$ is a hitting set of $G$
(otherwise $M$ is not maximal), but it is not necessarily minimal. Let us now
consider all sets $S\subseteq H$ with the following two properties: $|S|\le
d-1$ and all edges $e\in F$ have an element in $V\setminus S$ (in other words,
$V\setminus S$ is a hitting set of $G$). For such a set $S$ and a vertex $u\in
V\setminus H$ we will say that $u$ is seen by $S$, and write $u\in B(S)$, if
there exists $e\in F$ such that $e\cap H=S$ and $u\in e$. Intuitively, what we
are trying to do here is find a set $S$ that we will \emph{not} place in our
hitting set. Vertices seen by $S$ are then vertices which are more likely to be
placeable in a minimal hitting set.

Let $B_i$ be the union of all $B(S)$ for sets $S$ with $|S|=i$. Since every
vertex appears in a hyperedge, all vertices of $V\setminus H$ are seen by a set
$S$, and therefore belong in some $B_i$. Therefore, the union of all $B_i$ has
size at least $|V\setminus H| \ge n-n^{1/d} = \Omega(n)$. The largest of these
sets, then,  has size at least $\frac{n-n^{1/d}}{d} = \Omega(n)$.  Consider
then the largest such set, which corresponds to sets $S$ with size $s$. There
are at most ${|H|\choose s} = O(n^{s/d})$ such sets $S$. Since all together
they see $\Omega(n)$ vertices of $V\setminus H$, one of them must see at least
$\Omega(n^{1-\frac{s}{d}})$ vertices. Call this set $S_m$. Consider now the
hypergraph induced by $S_m\cup B(S_m)$. If we delete the vertices of $S_m$ from
this hypergraph, we get a hypergraph where every hyperedge has at most $d-s$
vertices. By induction, we can in polynomial time find a minimal hitting set of
this hypergraph with at least
$\Omega((n^{1-\frac{s}{d}})^{\frac{1}{d-s}})=\Omega(n^{1/d})$ vertices. Call
this set $H'$.

We are now ready to build our solution. Start with the set $V\setminus (S_m\cup
B(S_m))$ and add to it the vertices of $H'$. First, this is a hitting set,
because any hyperedge not hit by $V\setminus (S_m\cup B(S_m))$ is induced by
$S_m \cup B(S_m)$, and $H'$ hits all such hyperedges. We now proceed to make
this set minimal by arbitrarily deleting redundant vertices. The crucial point
here is that no vertex of $H'$ is deleted, since this would contradict the
minimality of $H'$ as a hitting set of the hypergraph induced by $S_m\cup
B(S_m)$. Thus, the resulting solution has size $\Omega(n^{1/d})$.}
\begin{pf}\proofofTheoremmmhsalg
\end{pf}
\begin{theorem}\label{cud_apx}
\CUD is 4-approximable in polynomial time\SV{, 3-approximable  with a running time in  $O^*(1.0883^{\tau(G)})$} and 2-approximable in time $O^*(1.2738^{\tau(G)})$ or $O^*(1.2132^n)$. \end{theorem}
\SV{\begin{prf}
First, find a vertex cover $V'$ in $G$ using any 2-approximation algorithm, and define $S'=V\setminus V'$. Let $S$ be a maximal independent set containing $S'$. $V\setminus S$ is a vertex cover of size $|V\!\setminus\! S| \leq |V'|\leq 2 \tau(G)\leq 4 (n-\Gamma(G))$, by eq.~\ref{complupperdom}. 
Moreover, $S$ is maximal independent and hence minimal dominating set which makes $V\!\setminus\! S$ a feasible solution for \CUD with $|V\!\setminus\! S|\leq 4(n-\Gamma(G))$. 
 The claimed running time for the factor-2 approximation stems from the best parameterised and exact algorithms \MVC by \cite{CheKanXia2010} and \cite{KneLanRos2009}, the factor-3 approxmation from the parameterised approximation in \cite{BraFer2013}. 
\end{prf}}
\begin{pf}Given a graph $G$ on $n$ vertices, we first find a vertex cover $V'$ in $G$ using any 2-approximation algorithm, and define $S'=V\setminus V'$. Set $S'$ is an independent set and let $S$ be a maximal independent set containing  $S'$. The set $V\setminus S$ is a vertex cover of size $|V\setminus S| \leq |V'|\leq 2 \tau(G)\leq 4 (n-\Gamma(G))$, by eq.~\ref{complupperdom}. 
 Moreover, $V\setminus S$ is the complement of a maximal independent set which also makes it the complement of a minimal dominating set, so overall a feasible solution for \CUD with $|V\setminus S|\leq 4(n-\Gamma(G))$. 
 The claimed running time for the factor-2 approximation stems from the best parameterised and exact algorithms \MVC by \cite{CheKanXia2010} and \cite{KneLanRos2009}. 
\end{pf}
\LV{
We could also use some results from parameterised approximation. For instance,
by \cite{BraFer2013}, we can conclude:
\begin{corollary}\CUD is 3-approximable  with a running time of  $O^*(1.0883^{\tau(G)})$.
\end{corollary}
}
With the results shown in the next section, we can conclude that 
\CUD is also APX-complete.

\LV{\todo[inline]{To the approx. group: Can we improve the poly-time factor?}}

\section{Graphs of bounded degree}

For these classes of graphs, we have also some new results on classical complexity.

\subsection{Classical complexity and exact algorithms}

\UD has been shown to be NP-hard on planar graphs of maximum degree six in \cite{AboHLMR2014} before. Here, we are going to strengthen this result in two directions, for cubic graphs and for planar subcubic graphs.

\begin{theorem} \label{UDcubic}
\UD is NP-hard on  cubic graphs. 
\end{theorem}

\SV{\begin{proof} (Sketch)
We present a reduction from \MIS on cubic graphs. Let $G=(V,E)$ be the cubic input graph.\\ 
\begin{minipage}{.6\textwidth}
Build $G'$ from $G$ by replacing every $(u,v)\in E$ by a  construction introducing six new vertices, as shown on the right.
Any $IS\subset V$ is an independent set for $G$ if and only if $G'$ contains an upper dominating set  of cardinality $|IS|+3|E|$.
\end{minipage}
\begin{minipage}{.4\textwidth}
\begin{center}
\includegraphics[scale=0.5]{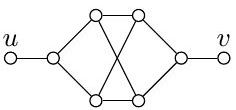}
\end{center}\hfill\framebox{}
\end{minipage}
\end{proof}}

\newcommand{\proofofTheoremUDcubic}{We present a 
reduction from \MIS on cubic graphs. Let $G=(V,E)$ be the cubic input graph for \MIS. Construct a cubic graph $G'$ from $G$ by replacing every edge $(u,v)\in E$ by the following construction introducing six new vertices for each edge:
\begin{center}
\includegraphics[scale=0.4]{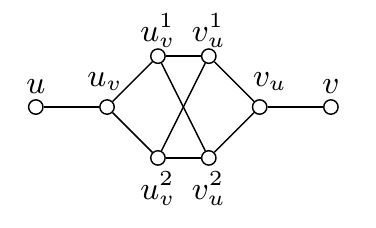}
\end{center}
  
If $IS$ is an independent set for $G$, the corresponding vertex-set in $G'$ can be extended to an upper dominating set $S$ of cardinality at least $|IS|+3|E|$ in the following way: For every edge $(u,v)$ with $v\notin IS$ add $\{v_u,u_v^1,u_v^2\}$  to $S$:

\begin{center}
\includegraphics[scale=0.4]{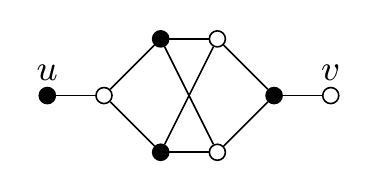}
\end{center}
Since $IS$ is independent, this procedure chooses three vertices for each edge-gadget in $G'$ and creates an independent set $S$ of cardinality $|IS|+3m$.  
If there was some vertex  not dominated by $S$ we could add it to $S$ without violating independence and $S$ will only increase in cardinality. Finally we arrive at a maximal independent and consequently minimal dominating set of cardinality at least $|IS|+3|E|$.

Let $S$ be an upper dominating set for $G'$. Minimality yields that for every edge $(u,v)\in E$ at most three of the vertices added for this edge can be in $S$. Consider an original edge $(u,v)\in E$ such that $u,v\in S$. From the vertices added for this edge, only two can be in $S$: Observe that any two out of  $\{u_v^1,u_v^2,v_u^1,v_u^2\}$  already dominate the whole subgraph. The set $\{u_v,v_u\}$ is also already dominating so there is no minimal dominating set of cardinality three in this case.

Consider the set $S'=S\cap V$ as potentially  independent set for the original graph $G$. If there are two vertices $u,v\in S'$ such that $(u,v)\in E$, the corresponding edge-gadget only adds two vertices to $S$. By successively deleting vertices from $S'$ as long as there is a conflict with respect to independence, we arrive at an independent set of cardinality at least $|S|-3|E|$.}
\begin{pf}\proofofTheoremUDcubic
\end{pf}

We complement this result by some results on exact algorithms.
Let us recall one important result on the pathwidth of subcubic graphs from~\cite{FomHoi2006}.

\begin{theorem}Let $\epsilon>0$ be given. 
For any subcubic graph $G$ of order $n>n_\epsilon$,
a path decomposition proving \LV{that }$pw(G)\leq n/6+\epsilon$ is computable in polynomial time. 
\end{theorem}

\LV{This result immediately gives an $O^*(1.2010^n)$-algorithm for solving \MD on subcubic graphs. We will take a similar route}\SV{We will use this result} to
prove moderately exponential-time algorithms for \UD on subcubic graphs.

\begin{proposition}\label{prop-UDpw}
\UD on graphs of pathwidth $p$ can be solved in time $O^*(7^p)$, given a corresponding path decomposition. 
\end{proposition}

\newcommand{\proofofPropUDpw}{We are considering all partitions of each bag of the path decomposition into 6 sets: $F$, $F^*$, $I$, $P$, $O$, $O^*$, where 
\begin{itemize}
\item $F$ is the set of vertices that belong to the upper dominating set and have already been matched to a private neighbour;
\item $F^*$  is the set of vertices that belong to the upper dominating set and still need to be  matched to a private neighbour;
\item $I$ is the set of vertices that belong to the upper dominating set and is independent in the graph induced by the upper dominating set;
\item $P$ is the set of private neighbours that are already matched to vertices in the upper dominating set;
\item $O$ is the set of vertices that are not belonging neither to the upper dominating set nor to the set of private neighbours but are already dominated;
\item $O^*$ is the set of vertices not belonging to the upper dominating set that have not been dominated yet.
\end{itemize}
(Sets within the partition can be also empty.) 
For each such partition, we determine the largest minimal dominating set in the situation described by the partition, assuming optimal settings in the part of the graph already forgotten.

We can assume that we are given a nice path decomposition. So, we only have to describe the table  initialisation (the situation in a bag containing only one vertex) and the table updates necessary when we introduce a new vertex into a bag and when we finally forget a vertex.

\begin{description}
\item[initialisation] We have six cases to consider:
\begin{itemize}
\item $T[\{v\},\emptyset,\emptyset,\emptyset,\emptyset,\emptyset]\gets -1$,
\item $T[\emptyset,\{v\},\emptyset,\emptyset,\emptyset,\emptyset]\gets 1$,
\item $T[\emptyset,\emptyset,\{v\},\emptyset,\emptyset,\emptyset]\gets 1$,
\item $T[\emptyset,\emptyset,\emptyset,\{v\},\emptyset,\emptyset]\gets -1$,
\item $T[\emptyset,\emptyset,\emptyset,\emptyset,\{v\},\emptyset]\gets -1$.
\item $T[\emptyset,\emptyset,\emptyset,\emptyset,\emptyset,\{v\}]\gets 1$.
\end{itemize}
Here, $-1$ signals the error cases when we try to introduce already dominated vertices.
\item[forget] Assume that we want to update table $T$ to table $T'$ for the partition $F$, $F^*$, $I$, $P$, $O$, $O^*$, eliminating vertex $v$:
\begin{itemize}
\item $T'[F\setminus\{v\},F^*,I,P,O,O^*]\gets T[F,F^*,I,P,O,O^*]$,
\item $T'[F,F^*\setminus\{v\},I,P,O,O^*]\gets -1$,
\item $T'[F,F^*,I\setminus\{v\},P,O,O^*]\gets T[F,F^*,I,P,O,O^*]$,
\item $T'[F,F^*,I,P\setminus\{v\},O,O^*]\gets T[F,F^*,I,P,O,O^*]$,
\item $T'[F,F^*,I,P,O\setminus\{v\},O^*]\gets T[F,F^*,I,P,O,O^*]$,
\item $T'[F,F^*,I,P,O,O^*\setminus\{v\}]\gets -1$.
\end{itemize}
Clearly, it is not feasible to eliminate vertices whose promises have not yet been fulfilled.
\item[introduce] We are now introducing a new vertex $v$ into the bag. The neighbourhood $N$ refers to the situation in the new bag, i.e., to  the corresponding induced graph. $T'$ is the new table and $T$ the old one.
\begin{itemize}
\item  $T'[F\cup\{v\},F^*,I,P,O,O^*]\gets -1$ if 
$N(v)\cap (I\cup O^*)\neq \emptyset$ or $|N(v)\cap P|\not=1$;\\
$T'[F\cup\{v\},F^*,I,P,O,O^*]\gets \max\{ T[F,F^*,I,P\setminus\{x\},O\setminus X,O^*\cup X\cup\{x\}]: x\in N(v), X\subseteq (N(v)\setminus\{x\})\cap O\}+1$ otherwise;\\
this means that exactly one neighbour $x$ of $v$ that was previously labelled  to be dominated in the future is selected as a private neighbour of $v$; all other neighbours of $v$ are labelled dominated;
\item $T'[F,F^*\cup\{v\},I,P,O,O^*]\gets -1$ if 
$N(v)\cap (I\cup P \cup O^*)\neq \emptyset$;\\
$T'[F,F^*\cup\{v\},I,P,O,O^*]\gets\max\{ T[F,F^*,I,P,O\setminus X,O^*\cup X]:  X\subseteq N(v)\cap O\}+1$ otherwise;\\
in contrast to the previous situation, no private neighbour has been selected;
\item $T'[F,F^*,I\cup\{v\},P,O,O^*]\gets -1$ if 
$N(v)\cap (I\cup F\cup F^*\cup P\cup O^*)\neq \emptyset$;\\
$T'[F,F^*,I\cup\{v\},P,O,O^*]\gets\max\{ T[F,F^*,I,P,O\setminus X,O^*\cup X]: X\subseteq N(v)\cap O\}+1$  otherwise;
\item  $T'[F,F^*,I,P\cup\{v\},O,O^*]\gets -1$ if 
$N(v)\cap I\neq \emptyset$ or $|N(v)\cap F|\neq 1$;\\
$T'[F,F^*,I,P\cup\{v\},O,O^*]\gets T[F\setminus   N(v),F^*\cup (N(v)\cap F),I,P,O,O^*]$ otherwise;\\
this means that exactly one neighbour $x$ of $v$ that was previously labelled as dominating but looking for a private neighbour in the future is selected as pairing up with $v$; all other neighbours of $v$ are not in the dominating set;
\item $T'[F,F^*,I,P,O\cup\{v\},O^*]\gets  T[F,F^*,I,P,O,O^*]$ if 
$N(v)\cap (F\cup F^*\cup I)\neq\emptyset
$ and $T'[F,F^*,I,P,O\cup\{v\},O^*]\gets -1$ otherwise;
\item $T'[F,F^*,I,P,O,O^*\cup\{v\}]\gets  T[F,F^*,I,P,O,O^*]$ unless $N(v)\cap (F\cup F^*\cup I)\neq\emptyset
$; in that case, 
$T'[F,F^*,I,P,O,O^*\cup\{v\}]\gets -1$.
\end{itemize}
\end{description}
The formal induction proof showing the correctness of the algorithm is an easy standard exercise.
As to the running time, observe that we cycle only in one case potentially through all subsets of $O$,
so that the running time follows by applying the 
binomial formula:
$$\sum_{i=0}^{p}\binom{p}{i}5^i 2^{p-i} = 7^p\,.$$}
\begin{pf}\proofofPropUDpw
\end{pf}

\LV{Observe that t}\SV{T}he upper bound on the running time can be improved for graphs of a certain maximum degree to $O^*(6^p)$, so that we can conclude:

\begin{corollary}\label{ud_ex}
\UD on subcubic graphs of order $n$ can be solved in time 
$O^*(1.3481^n)$, using the same amount of space.
\end{corollary}

\subsection{Parameterised complexity}

In contrast to the case of general graphs, \UD turns out to be easy (in the sense of paramterised complexity) for graphs of bounded degree.

\begin{proposition}\label{ud_fpt}
Fix $\Delta>2$. 
\UD\SV{,}\LV{ is in FPT when} restricted to graphs of maximum degree $\Delta$\SV{,}\LV{.
More precisely, the problem} can be solved in time $O^*((\Delta+1)^{2k})$.
\end{proposition}

\LV{The statement of the proposition is of course also true for $\Delta\in\{0,1,2\}$, but then the problem is (trivially) solvable in polynomial time. In the following, we give an argument based on branching.} 

\begin{prf}
Consider the simple branching algorithm that branches on all at most
$\Delta+1$ possibilities to dominate a yet undominated vertex.
Once we have fixed a new vertex in the dominating set,
we let follow another branch (of  at most
$\Delta+1$ possibilities) to determine the private neighbour of the new vertex in the dominating set.
Assuming that we are only looking for sets of size $k$,
we can find a yes instance in each branch where we needed to put
$k$ vertices in the dominating set (so far); if that set is not yet dominating, we can turn it into a minimal dominating set by a greedy approach, respecting previous choices.
\LV{The overall running time of the branching algorithm
is hence $O^*((\Delta+1)^{2k})$.}
\end{prf}

The astute reader might wonder why we have to do this unusual
2-stage branching, but recall Theorem~\ref{thm-MDSE-hardness} that shows that we cannot extend some set of vertices of size at most $k$ that easily to a minimal dominating set containing it.

\LV{\todo[inline]{HF: It might be interesting to study MDSE, parameterised by the size of the input vertex subset $S$, even on degree-bounded graphs. A nice solution would immediately improve on the branching that we presented.}}


Brooks' Theorem yields the following result.

\begin{proposition}\label{prop-UDkernel}
Fix $\Delta>2$. 
\UD has a problem kernel with at most $\Delta k$ many vertices.
\end{proposition}

\newcommand{\proofofPropUDkernel}{First, we can assume that the input graph $G$ is
connected, as otherwise we can apply the following argument 
separately on each connected component.
Assume $G$ is a cycle or a clique. Then, the problem \UD can be optimally solved in polynomial time, i.e., we can produce a kernel as
small as we want.
Otherwise, Brooks' Theorem yields a polynomial-time algorithm
that produces a proper colouring of $G$ with (at most) $\Delta$ many colours. Extend the biggest colour class to a maximal independent set $I$  of $G$. As $I$ is maximal, it is also a minimal dominating set.
So, there is a minimal dominating set $I$ of size at least $n/\Delta$, where
$n$ is the order of $G$. So, $\Gamma(G)\geq n/\Delta$.
If $k<n/\Delta$, we can therefore immediately answer YES.
In the other case, $n\leq \Delta k$ as claimed.}
\begin{pf}\proofofPropUDkernel
\end{pf}

With some more combinatorial effort, we obtain:

\begin{proposition}\label{prop-CUDkernel}
Fix $\Delta>2$. \CUD has a problem kernel with at most $(\Delta+0.5)\ell$ many vertices.
\end{proposition}

 \newcommand{\proofofPropCUDkernel}{Consider any graph $G=(V,E)$. For any partition $(F,I,P,O)$ corresponding to an upper dominating set $D=I\cup F$ for $G$, isolated vertices in $G$ always belong to $I$ and can hence be deleted in any instance of \CUD without changing $\ell$. For any graph $G$ without isolated vertices, the set $P\cup O$ is a dominating set for $G$, since $\emptyset \not=N(v)\subset O$ for all $v\in I$ and $N(v)\cap P\not=\emptyset $ for all $v\in F$. Maximum degree $\Delta $ hence immediately implies $n=|N[P\cup O]|\leq (\Delta+1)\ell$. 

Since any connected component can be solved separately, we can assume that $G$ is connected. For any $v\in P$, the structure of the partition $(F,I,P,O)$ yields $|N[v]\cap D|=1$, so either $|N[v]|=1<\Delta$ or there is at least one $w\in P\cup O$ such that $N[v]\cap N[w]\not=\emptyset$. For any $v\in O$, if $N[v]\cap F\not=\emptyset$, the $F$-vertex in this intersection has a neighbour $w\in P$, which means $N[w]\cap N[v]\not=\emptyset$. If $N[v]\subset I$ and $N[v]\not=V$, at least one of the $I$-vertices in  $N[v]$ has to have another neighbour to connect to the rest of the graph. Since $N[I]\subset O$, this also implies the existence of a vertex $w\in O$, $w\not=v$ with $N[w]\cap N[v]\not=\emptyset$. Finally, if $N[v]\not\subset I\cup F$, there is obviously a $w\in P\cup O$, $w\not=v$ with $N[w]\cap N[v]\not=\emptyset$.

Assume that there is an upper dominating set with partition $(F,I,P,O)$ such that $|P\cup O|=l\leq \ell$ and let $v_1,\dots,v_l$ be the $l>1$ vertices in $P\cup O$. By the above argued domination-property of $P\cup O$, we have:
$$ n= |\bigcup _{i=1}^l N[v_i]|=\tfrac12\sum_{i=1}^l |N[v_i]\setminus\bigcup_{j=1}^{i-1}N[v_j]|+ \tfrac12\sum_{i=1}^l |N[v_i]\setminus\!\!\bigcup_{j=i+1}^{l}\!\!N[v_j]|$$
Further, by the above argument about neighbourhoods of vertices in $P\cup O$, maximum degree $\Delta$ yields  for every $i\in \{1,\dots,l\}$ either  $|N[v_i]\setminus\bigcup_{j=1}^{i-1}N[v_j]|\leq \Delta$ or $|N[v_i]\setminus\bigcup_{j=i+1}^lN[v_j]|\leq \Delta$ which gives:  $$n=\tfrac12 \sum_{i=1}^l |N[v_i]\setminus\bigcup_{j=1}^{i-1}N[v_j]| + |N[v_i]\setminus\bigcup_{j=i+1}^lN[v_j]|\leq \tfrac12 l(2\Delta +1)\leq(\Delta+0.5)\ell.$$
Any graph with more than $(\Delta+0.5)\ell$ vertices is consequently a NO-instance which yields the stated kernelisation, as the excluded case $|P\cup O|=1$ (or in other words $N[v]=V$ for some $v\in O$) can be solved trivially.}
\begin{pf}\proofofPropCUDkernel
\end{pf}


This implies that we have a $3k$-size vertex kernel for \UD, restricted to subcubic graphs, and a $3.5\ell$-size vertex kernel for \CUD,  again
restricted to subcubic graphs.
With \cite[Theorem 3.1]{Cheetal2007}, we can conclude the following consequence:

\begin{corollary}
Unless $P$ equals $NP$, for any $\varepsilon>0$, \UD, restricted to subcubic graphs, does not admit a kernel with less than $(1.4-\varepsilon)k$ vertices; neither does \CUD, restricted to subcubic graphs, admit a kernel with less than $(1.5-\varepsilon) \ell$ vertices.
\end{corollary}

\subsection{Approximation}

Using that \MIS on cubic graphs is APX-hard \cite{AliKan2000}, one can however obtain the following result from the proof of Theorem~\ref{UDcubic}. 

\begin{theorem}\label{ud_ptas}
\UD is APX-hard even for cubic graphs. 
\end{theorem}

\begin{pf} 
The reduction from \MIS on cubic graphs in  the proof of Theorem~\ref{UDcubic} is an $L$-reduction. Given a graph $G$ on $n$ vertices and $m$ edges, as an instance of \MIS, we construct a graph $G'$ instance of \UD with the following properties: $\opt(G')=\opt(G)+3m$ and given any solution of size $val'$ in $G'$, we can construct a solution of size $val=val'-3m$. 
Thus $\opt(G') \leq 19 \opt(G)$ since $\opt(G) \geq n/4$.  Moreover, $\opt(G)-val=\opt(G')-val'$. 
\end{pf}

We can obtain membership in APX, so altogether APX-completeness,
\LV{by re-considering the simple greedy algorithm that was the basis 
of the kernel argument given above.}\SV{by the simple fact that any dominating set has cardinality at least $\frac {n}{\Delta+1}$ for graphs of maximum degree $\Delta$.}
However, we can do much better, as we now exhibit\LV{ in the following theorem}.

\begin{theorem}\label{ud_delta_approx}
Consider some graph-class~$\mathcal{G}(p,{\rho})$ with the following properties:\begin{itemize}
\item 
One can colour every $G \in \mathcal{G}(p,{\rho})$ with~$p$ colours in polynomial time. 
\item 
For any $G \in \mathcal{G}(p,{\rho})$, \MIS is $\rho$-approximable in polynomial time. 
\end{itemize}
Then, in every $G \in \mathcal{G}(p,{\rho})$ \UD is approximable in polynomial time within ratio at most:
\begin{equation}\label{ratio}
\max\left\{\rho, \frac{\Delta\rho p + \Delta - 1}{2\rho\Delta}\right\}
\end{equation}
\end{theorem}

The basic proof idea  is based upon equation (\ref{upperdomdelta_is}) and the fact that any maximal  independent set is a feasible \UD-solution. Then, we distinguish two cases, and we run two \MIS-algorithms efficiently handling them. We finally output the best among the computed solutions.

Any graph of maximum degree~$\Delta$ can be coloured with at most~$\Delta$ colours~\cite{Lov75}; furthermore, \MIS is approximable within ratio $(\Delta+3)/5$ in graphs of maximum degree~$\Delta$~\cite{BerFuj95}. So, the class~$\mathcal{G}(\Delta,(\Delta+3)/5)$ contains all  graphs of maximum degree $\Delta$. Conversely,~$\mathcal{G}(4,1+\epsilon)$ contains the planar graphs, and~$\mathcal{G}(3,1+\epsilon)$ the triangle-free planar graphs, since \MIS admits a PTAS for such graphs~\cite{Bak94}.  Hence, from Theorem~\ref{ud_delta_approx}, the following corollary can be derived.
\begin{corollary}\label{ud_delta_all}
\UD is approximable in polynomial time within ratios $(6\Delta^2+2\Delta-3)/10\Delta$ in general graphs,
%
%
$5/2 + \epsilon$ in planar graphs and $2 + \epsilon$ in triangle-free planar graphs, for any $\epsilon > 0$.
\end{corollary}

Let us note that \LV{ the result of} Theorem~\ref{ud_delta_approx} can be easily improved in the case of regular graphs. Indeed, in this case
$\Gamma(G) \leqslant \frac{n}{2}$ according to~\cite{HenSla96}.
Then\LV{, from the proof of Theorem~\ref{ud_delta_approx}} one can conclude\SV{:} 
\LV{ the following corollary.}

\begin{corollary}\label{ud_delta_approx_cor}
\UD in regular graphs is approximable in polynomial time within ratio~$\Delta/2$.
\end{corollary}
We are now turning to the complementary problem, i.e., to \CUD. Notice that we know that this problem lies in APX for general graphs, more precisely it is 4-approximable in polynomial time. 
Since \MVC  is 7/6-approximable in polynomial time~\cite{BerFuj95} for cubic graphs, using  the same argument as in the proof of Theorem~\ref{cud_apx}, we obtain a polynomial-time 7/3-approximation  for \CUD on cubic graphs.\\


\begin{theorem}\label{cud_ptas}
\CUD is APX-complete even for cubic graphs. 
\end{theorem}\SV{This follows from the proof of Theorem~\ref{UDcubic} which can  be seen as an L-reduction from \MIS to \CUD.}
\newcommand{\proofofcudptas}{We re-consider the same reduction from \MIS on cubic graphs as in the proof of Theorem~\ref{UDcubic} and prove that it is indeed an $L$-reduction also for this case. Given a graph on $n$ vertices and $m$ edges, as an  instance of \MIS, we construct a graph $G'$, an instance of \CUD on $n'$ vertices with the following properties: $\opt(G')=n'-\opt(G)-m$ and given any solution of size $val'$ in $G'$, we can construct a solution of size $val=n'-3m-val$. 
Thus $\opt(G') \leq c n- n/4-9n/2=c' n\leq \opt(G)$, since $\opt(G) \geq n/4$.  Moreover, $\opt(G)-val=\opt(G')-val'$.
}
\LV{\begin{pf}
\proofofcudptas
\end{pf}
}

\LV{\section{Planar graphs}

Notice that the edge gadget that we used in the proof of Theorem~\ref{UDcubic} destroys planarity, so this does not show that \UD is NP-hard on planar cubic graphs, as it is known for \MIS.
However, as omitting the two crossing edges of the gadget (as the reader can verify), we can 
obtain the following result immediately:

\begin{corollary}\label{UDcubicplanar}
\UD is NP-hard on  subcubic planar graphs. 
\end{corollary}

\todo[inline]{To the approx. group: Are there things in your NP-hardness proof of this result that are worth maintaining?! Then, please insert this here. I know that the one-paragraph note of ours is pretty short.}

\todo[inline]{Mathieu, can you see into technical details if / how Corollary~\ref{ud_ex} extends to tree decompositions, as this is important to know. Of course we believe that such a thing exists, but 
does it also have 6 in the basis of the running time?}



Similar constructions should be possible for graphs of any bounded genus.

Let us finally mention that some further tricks should be possible to improve on the running times or other practical aspects both of subexponential-type algorithms and of PTAS for planar graphs.
We only refer to the discussions in \cite{MarGuJia2009,MarGu2013} and the literature quoted therein with respect to practical aspects of the computation of $\gamma(G)$ for planar graphs.}



\section{Discussions and open problems}

The motivation to study \UD (at least for some in the group of authors)
was based on the following observation:

\begin{proposition}
\UD can be solved in time $O^*(1.7159^n)$ on general graphs of order $n$.
\end{proposition}

\begin{pf}
The suggested algorithm simply lists all minimal dominating sets and then picks the biggest one.
It has been shown in \cite{Fometal2008a} that this 
enumeration problem can be performed in the claimed running time.
\end{pf}

It is of course a bit nagging that there seems to be no better algorithm (analysis) than this  enumeration algorithm for \UD. Recall that the minimisation counterpart can be solved in better than 
$O^*(1.5^n)$ time \cite{Iwa1112,RooBod2011}. 
Intuitively, the problem behind finding nothing better than 
enumeration has to do with the hardness of the extension problem
considered in Section~\ref{sec-extension}, as it is hard to fill up
partial solutions determined by branching.
As this appears to be quite a tough problem, it makes a lot of sense to study it on restricted graph classes. This is what we did
above for subcubic graphs, see Corollary~\ref{ud_ex}.
We are currently working on \UD on planar graphs, which turns out to be a bit harder than with other graph problems, again due to the hardness of the 
extension problem.
We summarise some open problems.

\begin{itemize}

\item Is \UD in W[1]?


\item Can we improve on the \LV{current} 4-approximation of \CUD? 

\LV{\todo[inline]{To the approx. group: Can you solve this?}}

\item Can we find smaller kernels for \UD or \CUD on degree-bounded graphs?

\item Can we find  exact (e.g., branching) algorithms that beat enumeration
or pathwidth-based algorithms for \UD, at least on cubic graphs?
\end{itemize}

\paragraph{Acknowledgements.}
We like to thank our colleagues 
David Manlove and 
Daniel Meister for some discussions on upper domination.
Part of this research was supported by the Deutsche Forschungsgemeinschaft, grant FE 560/6-1.

\bibliographystyle{plain}

\bibliography{abbrev,hen}

\SV{
\newpage
\section{Appendix: Omitted (standard) definitions}

\subsection{Basic notions of graph theory}\label{basic_notions}
We use $G=(V,E)$ to denote a simple undirected  graph $G$ where $V$ is the set of vertices of $G$ and $E$ the set of edges of $G$. 
The number of vertices $|V|$ is also known as the order of $G$. 
As usual, $N(v)$ denotes the open neighbourhood of $v$ in a graph $G$, and $N[v]$ is the closed neighbourhood of $v$ in $G$, i.e., $N[v]=N(v)\cup\{v\}$. 
We extend these notions to vertex sets, e.g., for $X \subseteq V$ we have $N(X)=\bigcup_{x\in X}N(x)$ and $N[X]=\bigcup_{x\in X}N[x]$.

The cardinality of $N(v)$ is also known as the degree of $v$,
denoted as $deg(v)$. The maximum degree in a graph is usually written
as $\Delta$. A graph of maximum degree three is called subcubic, and if  all degrees equal three, it is called a cubic graph.

 Given a  graph $G=(V,E)$, a subset $S$ of $V$ is a \emph{dominating set} if every vertex $v\in V\setminus S$ has at least one neighbour in $S$, i.e., if $N[S]=V$. A dominating set is minimal if no proper subset of it is a dominating set.
 Likewise, a vertex set $I$ is \emph{independent} if $N(I)\cap I=\emptyset$. An independent set is maximal if no proper superset of it
 is independent. 
 
 \subsection{$E$-reductions}
  A
problem $A$ is called {\it $E$-reducible} to a problem $B$,
if there exist polynomial 
time computable functions $f$, $g$ and a
constant $\beta$ such that
\begin{itemize}
\item $f$ maps an instance $I$ of $A$ to an instance $I'$ of $B$
such that $\opt(I)$ and $\opt(I')$ are related by a polynomial
factor, i.e. there exists a polynomial $p$ such that
$\opt(I')\leq p(|I|) \opt(I)$, \item $g$ maps any solution $S'$ of $I'$
to one solution $S$ of $I$ such that $\varepsilon(I,S)\leq \beta
\varepsilon(I',S')$.
\end{itemize}

An important property of an  $E$-reduction is that it can be applied
uniformly to all levels of approximability; that is, if $A$ is
$E$-reducible to $B$ and $B$ belongs to $\cal{C}$ then $A$
belongs to $\cal{C}$ as well, where $\cal{C}$ is a class of
optimisation problems with any kind of approximation guarantee.\footnote{See also 
S.~Khanna, R.~Motwani, M.~Sudan, and U.~Vazirani.
 On syntactic versus computational views of approximability.
 {\em {SIAM} Journal on Computing}, 28:164--191, 1998.}
 
 \section{Omitted Proofs}
 
 In this section, we collect proofs that have been omitted or considerably shortened in the main text body.
 
 \subsection{Proof of Lemma~\ref{|I|<=alpha-2}}
 \proofofLemmaalphatwo
 
 \subsection{Proof of Lemma~\ref{bounds_on_Gama}}
 \proofofLemmaboundsonGama
 
 \subsection{Proof of Lemma~\ref{bounds_on_Gama_with_Delta}}
\proofofLemmaboundsonGamawithDelta

 \subsection{Proof of Theorem~\ref{thm-MDSE-hardness}}
\proofofTheoremMDSEhardness

\subsection{Proof of Theorem~\ref{w_hardness}}
\proofofTheoremWhardness

\subsection{Proof of Proposition~\ref{prop-Wtwo}}
\proofofPropWtwo

\subsection{Proof of Theorem~\ref{cud_kernel}}
\proofofTheoremcudkernel

\subsection{Details on Proposition~\ref{prop-CoUDbranching}}
\CoUDbranching

\subsection{Proof of Theorem~\ref{mmhs}}
\proofofTheoremmmhs

\subsection{Proof of Cor.~\ref{mmhs_apx}}
\proofofCormmhsapx

\subsection{Proof of Theorem~\ref{ud_apx}}
\proofofTheoremudapx

\subsection{Proof of Theorem~\ref{mmhs_alg}}
\proofofTheoremmmhsalg


\subsection{Proof of Theorem~\ref{UDcubic}}
\proofofTheoremUDcubic

\subsection{Proof of Proposition~\ref{prop-UDpw}}
\proofofPropUDpw{}

\subsection{Proof of Proposition~\ref{prop-UDkernel}}

\proofofPropUDkernel

\subsection{Proof of Proposition~\ref{prop-CUDkernel}}
\proofofPropCUDkernel

\subsection{Proof of Theorem~\ref{ud_delta_approx}}

The approximation algorithm consists of running two independent set algorithms, by greedily augmenting solutions computed in order to become maximal for inclusion and of returning the best among them, denoted by~$U$. Recall that any maximal (for inclusion) independent set is a feasible upper dominating set.

The algorithms used are:
\begin{itemize}
\item[(i)] the $\rho$-approximation algorithm assumed for~$\mathcal{G}(p,{\rho})$ and
\item[(ii)] the (also assumed) algorithm that colours the vertices of the input graph with~$p$ colours and takes the largest colour as solution for \UD.
\end{itemize}
Revisit~(\ref{upperdomdelta_is}). If the maximum there, is realised by the first term, then we are done since the $\rho$-approximation \MIS-algorithm assumed, also achieves ratio~$\rho$ for \UD.

Suppose now that the maximum in~(\ref{upperdomdelta_is}) is realised by the second term and set $\alpha(G) = n/t$, for some $t \geqslant 1$ that will be fixed later. In order to simplify calculations we will use the following bounds for~$\Gamma(G)$, easily derived from~\ref{upperdomdelta_is}:
\begin{eqnarray}\label{upperdomdelta_is_w}
\Gamma(G) &\leqslant& \frac{n}{2} + \frac{\alpha(G)(\Delta-1)}{2\Delta} = \frac{\Delta(t+1)-1}{2t\Delta}n \label{upperdomdelta_is_w1} \\
\Gamma(G) &\leqslant& \frac{\Delta(t+1) -1}{2\Delta}\alpha(G) \label{upperdomdelta_is_w2}
\end{eqnarray}
Expression~(\ref{upperdomdelta_is_w2}) leads to:
\begin{equation}\label{upperdomdelta_is_w21}
\alpha(G) \geqslant \frac{2\Delta}{\Delta(t+1)-1}\Gamma(G)
\end{equation}
If the solution~$U$ returned by the algorithm is the one of~item~(i), then, using~(\ref{upperdomdelta_is_w21}), it holds that:
\begin{align}\label{ratio1}
|U| \geqslant & \rho\alpha(G) \geqslant \frac{2\rho\Delta}{\Delta(t+1)-1}\Gamma(G) \nonumber \\
\frac{\Gamma(G)}{|U|} \leqslant & \frac{\Delta(t+1)-1}{2\rho\Delta}
\end{align}
Assume now that the \UD-solution returned is the one by item~(ii) and note that the largest colour computed is assigned to at least~$n/p$ vertices of the input graph which, obviously, form an independent set. So, in this case, $|U| \geqslant n/p$ and using~(\ref{upperdomdelta_is_w1}), the ratio achieved is:
\begin{equation}\label{ratio2}
\frac{\Gamma(G)}{|U|} \leqslant \frac{\frac{\Delta(t+1)-1}{2t\Delta}n}{\frac{n}{p}} = \frac{p(\Delta(t+1)-1)}{2t\Delta}
\end{equation}
Equality of ratios in~(\ref{ratio1}) and~(\ref{ratio2}) implies $t = \rho p$ and setting it to either one of those leads to the second term of the $\max$ expression in~(\ref{ratio}).

\subsection{Proof of Theorem~\ref{cud_ptas}}
\proofofcudptas
}

\end{document}